\documentclass[10pt, draftclsnofoot, twocolumn]{IEEEtran}
\pdfoutput=1
\usepackage{amssymb}
\usepackage{algorithm}
\usepackage{algpseudocode}
\usepackage{algorithmicx}
\usepackage{amsfonts}
\usepackage{amsmath}
\usepackage{amssymb}
\usepackage{amsthm}
\usepackage{array}
\usepackage{bbding}
\usepackage{bigints}
\usepackage{booktabs}
\usepackage{cite}
\usepackage{geometry}
\usepackage{color}
\usepackage{diagbox}
\usepackage{epsfig,latexsym}
\usepackage{epstopdf}
\usepackage{graphicx}
\usepackage{fancyhdr}
\usepackage{float}
\usepackage{indentfirst}
\usepackage{lastpage}
\usepackage{makecell}
\usepackage{mathtools}
\usepackage{multirow}
\usepackage{pifont}
\usepackage{psfrag}
\usepackage{setspace}
\usepackage{stfloats}
\usepackage[colorlinks,linkcolor=blue]{hyperref}
\hypersetup{hidelinks,
	colorlinks=true,
	allcolors=black,
	pdfstartview=Fit,
	breaklinks=true}
\usepackage{cleveref}
\renewcommand{\baselinestretch}{1.0}
\usepackage{empheq}
\usepackage{enumerate}
\usepackage{times}
\usepackage{subfigure}
\renewcommand{\arraystretch}{1}
\newcolumntype{C}{>{\centering\arraybackslash$}p{\linewidth}<{$}}

\newtheorem{theorem}{Theorem}

\newtheorem{lemma}{Lemma}

\newtheorem{corollary}{Corollary}

\newcommand{\RNum}[1]{\uppercase\expandafter{\romannumeral #1\relax}}

\begin{document}
\title{Multi-bit Distributed Detection of Sparse Stochastic Signals over Error-Prone Reporting Channels}

\author{{Linlin Mao}, {\em Member,~IEEE}, {Shefeng Yan}, {\em Senior Member,~IEEE}, {Zeping Sui}, {\em Member,~IEEE},\\ and {Hongbin Li}, {\em Fellow,~IEEE}
\thanks{This work was supported in part by the National Natural Science Foundation of China under Grants 62371447 and 62192711. The work of Hongbin Li was supported in part by the National Science Foundation under Grants ECCS-1923739, ECCS-2212940, ECCS-2332534, and CCF-2316865. (\emph{Corresponding author: Shefeng Yan})}
\thanks{Linlin Mao and Shefeng Yan are with the Institute of Acoustics, Chinese Academy of Sciences, Beijing 100190, China (e-mail: maoll@mail.ioa.ac.cn;sfyan@ieee.org).}
\thanks{Zeping Sui is with the School of Computer Science and Electronics Engineering, University of Essex, Colchester CO4 3SQ, U.K. (e-mail: zepingsui@outlook.com).}
\thanks{Hongbin Li is with the Department of Electrical and Computer Engineering, Stevens Institute of Technology, Hoboken, NJ 07030 USA (e-mail: hongbin.li@stevens.edu).}
}
\maketitle

\begin{abstract}
We consider a distributed detection problem within a wireless sensor network (WSN), where a substantial number of sensors cooperate to detect the existence of sparse stochastic signals. To achieve a trade-off between detection performance and system constraints, multi-bit quantizers are employed at local sensors. Then, two quantization strategies, namely raw quantization (RQ) and likelihood ratio quantization (LQ), are examined. The multi-bit quantized signals undergo encoding into binary codewords and are subsequently transmitted to the fusion center via error-prone reporting channels. Upon exploiting the locally most powerful test (LMPT) strategy, we devise  two multi-bit LMPT detectors in which quantized  raw observations and local likelihood ratios are fused respectively. Moreover, the asymptotic detection performance of the proposed quantized detectors is analyzed, and closed-form expressions for the detection and false alarm probabilities are derived. Furthermore, the multi-bit quantizer design criterion, considering both RQ and LQ, is then proposed to achieve near-optimal asymptotic performance for our proposed detectors. The normalized Fisher information and asymptotic relative efficiency are derived, serving as tools  to analyze and compensate for the loss of information introduced by the quantization. Simulation results validate the effectiveness of the proposed detectors, especially in scenarios with low signal-to-noise ratios and poor channel conditions.
\end{abstract}
\begin{IEEEkeywords}
Wireless sensor networks, distributed detection, sparse signal, multi-bit quantizer, error-prone channels.
\end{IEEEkeywords}
\IEEEpeerreviewmaketitle
\vspace{-0.9em}
\section{Introduction}\label{Section1}
Wireless sensor networks (WSNs) have gained considerable interest in recent years due to their flexibility and exceptional performance in several activities such as data collecting~\cite{wei2021reliable,cheng2022dynamic}, distributed detection~\cite{aldalahmeh2019fusion}, signal estimation~\cite{van2021distributed}, vehicle tracking~\cite{liang2020distributed}, and target localization~\cite{yan2018feedback}. Many WSN applications, such as disaster prevention, appliance monitoring, and battlefield surveillance, naturally face the challenges of distributed detection~\cite{braca2015distributed,maya2021fully,zhu2022adaptive,tabella2023time}. Notably, copula-based fusion models were exploited  to attain  superior performance in multi-sensor remote sensing and target detection~\cite{wang2022proposal,li2024comic}.
A distributed detection system is usually composed of multiple geographically distributed sensor nodes that make noisy observations over a region of interest. These nodes then transmit their  processed data to a fusion center (FC), which aggregates the received data to obtain a global decision.

One key characteristic of WSN is that the channel bandwidth is restricted and sensor nodes typically have limited power supplies~\cite{zhang2020Distributed}, imposing severe constraints on the amount of  communication data. To minimize communication data amount, a direct method is to convert the raw observations from local sensors into one-bit data, which we term as raw quantization (RQ). Several distributed detection schemes using one-bit RQ fusion have been studied, such as the  generalized likelihood ratio test (GLRT)~\cite{fang2013one}, Rao~\cite{ciuonzo2013one}, locally most powerful test (LMPT)~\cite{wang2019spl}, generalized variants of Rao~\cite{ciuonzo2017generalized} and locally optimum detection (LOD) tests~\cite{ciuonzo2021distributed}, sequential test~\cite{hu2018decentralized}, as well as joint detection and estimation methods \cite{zang2022fast}, respectively. Likelihood-ratio (LR) based quantization (LQ), an alternative to RQ fusion compute and quantize the LRs at local sensors before forwarding them to the FC. It is worth noting that hard decision fusion can be considered as an extreme case of transmitting quantized LRs~\cite{javadi2016detection}. For uncooperative targets with sensor-target distance dependent amplitude attenuation, decision fusion based on GLRT and generalized LOD have been investigated in \cite{ciuonzo2017distributed} and \cite{ciuonzo2017quantizer}, respectively, whereas decision fusion with adaptive topology was taken into consideration in \cite{luo2019distributed}. 

One-bit quantization-based distributed detection provides the advantages of minimal data transmission and low communication energy consumption; nevertheless,  in comparison to centralized detection, a significant amount of information is lost,  leading to a loss of detection performance. To bridge the performance gap,  researchers have looked into multi-bit quantization techniques~\cite{wang2021target,hu2020decentralized,gao2014quantizer,cheng2019multibit,cheng2021multi,zhang2023direct,yang2023hybrid}. The main challenge with these techniques is high complexity, which arises from a nonlinear multi-dimensional search procedure required by these methods. In particular, based on multi-bit uniformly quantized data, the distributed signal detection problem was examined  for cloud MIMO radar \cite{wang2021target} and, respectively, underwater sensing based on a reflection model with aspect and distance-dependent received signals \cite{hu2020decentralized}. The quantification step in the former can be easily calculated given the fixed quantization interval and quantization depth, whereas in the latter it is optimized by maximizing the Kullback-Leibler divergence between the null and alternative hypotheses. On the other hand, non-uniform multi-bit RQ fusion has been employed to detect an unknown signal parameter in three different contexts: Gaussian noise \cite{gao2014quantizer}, zero-mean unimodal symmetric noise \cite{cheng2019multibit}, and a model accounting for distance-dependent signal amplitude decay \cite{cheng2021multi}. Particle swarm optimization  algorithms (PSOA)  were employed by all three to optimize multi-bit quantization threshold designs.

In addition to quantization, compression is another efficient method for lowering the transmission load. To meet the bandwidth constraint, both compression and quantization are performed  in \cite{ciuonzo2020bandwidth} for the decentralized detection of an unknown vector signal using GLRT and Rao tests without the sparse detected signal assumption. In many applications, the signal to be detected is usually sparse or has a sparse representation in a  transformed domain. Consequently, compressive sensing (CS) is capable of attaining compression without significant  performance loss~\cite{donoho2006compressed}. Distributed detection of an unknown deterministic sparse vector signal based on one-bit CS measurements was considered in \cite{zayyani2016double}, where sparse recovery was employed to reconstruct the detection statistic. Research on the distributed detection of sparse stochastic signals without  sparse recovery was carried out in \cite{li2019secure,li2020censoring,li2020falsified,wang2018detection,feng2023compressive,wang2019detection,wang2019gg,li2020tree,li2019lr,Mohammadi2022generalized}. In  \cite{li2020censoring} and  \cite{li2020falsified}, distributed detection based on compressive measures was integrated with censoring and falsified censoring techniques to address local energy supply restrictions with improved system secrecy, respectively. The LMPT, Rao and Wald tests for compressive detection of sparse stochastic signals with unknown sparsity degree were explored in \cite{wang2018detection} and \cite{feng2023compressive}, which we shall refer to as Clairvoyant detectors in the following. Apart from Clairvoyant detectors, quantized detectors for distributed detection of sparse stochastic signals were also investigated in \cite{wang2019detection,wang2019gg,li2020tree,li2019lr,Mohammadi2022generalized}.  Multi-bit RQ fusion was employed in \cite{wang2019detection} and \cite{wang2019gg} to detect a sparse signal from Gaussian and generalized Gaussian noise, respectively. In \cite{li2020tree,li2019lr} and \cite{Mohammadi2022generalized}, one-bit LQ fusion was utilized to improve the detection performance of RQ detectors, whereby a non-ideal reporting channel was also considered in \cite{Mohammadi2022generalized}. Nevertheless, for the distributed detection of sparse stochastic signals over error-prone channels -- which is typically the practical case with restricted channel capacity -- how to use multi-bit LQ fusion and multi-bit RQ fusion has not yet been studied in the literature. Multi-bit quantization is required, as will be indicated in Section \ref{Section4}, particularly over imperfect reporting channels since one-bit quantization may result in a significant loss of Fisher information.

Against this background, we  design  multi-bit quantizers for distributed detection of a sparse stochastic signal over error-prone reporting channels. The contributions of our paper are explicitly contrasted with the existing literature in Table~\ref{table1}, which is further detailed below.
\begin{itemize}
	\item Multi-bit quantized detectors based on both raw quantization and LR quantization are proposed. To strike the detection performance \emph{vs.} system constraints trade-off, we conceive multi-bit quantization on both the raw observations and the LRs at each local sensor. The multi-bit quantized signals are encoded as binary codewords and then sent to the FC via error-prone reporting channels, which are modeled as binary symmetric channels (BSCs). We then develop two multi-bit LMPT detectors that can fuse raw observations and local LRs respectively. 
	\item We then analyze the asymptotic detection performance of the proposed quantized detectors, and derive closed-form expressions for the detection and false alarm probabilities. The results demonstrate that the asymptotic distribution of the LMPT test statistics follows a non-central Normal distribution, characterized by a non-central parameter that increases monotonically with the Fisher information (FI) of each detector, which depends on both the quantization threshold and the crossover probability of the BSC.
	\item We further derive the design criteria for multi-bit quantizer that consider both RQ and LQ, aiming to achieve near-optimal asymptotic performance for the proposed quantized detectors. To enhance the detection performance at the FC, we formulate an optimization problem concerning the FI to determine the optimum quantization thresholds. It should be noted that our proposed  criteria are not a trivial extension of \cite{gao2014quantizer}, since their signal models and detection strategies are different, and different quantization methods are invoked in the LQs. Moreover,  the design of multi-bit quantizers based on LQ fusion for both ideal and practical channels has not yet appeared in the literature. In this work, the ideal channel scenario can be regarded as a special case with a zero crossover probability. 
	\item The normalized Fisher information is derived to analyze the loss of information due to quantization. This provides a useful method for visually characterizing quantizers under varying channels. Additionally, an analytical derivation of the asymptotic relative efficiency (ARE) is provided to assess  the loss of information. The ARE can be employed to determined  the number of additional sensors that a quantized  detector requires relative to a Clairvoyant detector. Comparative  analysis between LQ and RQ is also provided to clarify why the latter performs worse than the former at low-bit quantization.
	\item Furthermore, we benchmark the performance of the proposed quantized detectors with Clairvoyant and  uniformly quantized solutions across a range of false alarm probabilities, number of sensors, and signal-to-noise ratios (SNR) under the conditions of different quantization depths and cross probabilities. It is observed that the proposed quantizers perform better than the conventional counterparts, especially in  low SNR and poor channel  scenarios.
	\end{itemize}
\begin{table*}[!htbp]
\footnotesize
\centering
\caption{Contrasting Our Contributions to the Existing Literature}
\label{table1}
\scalebox{1}{
\begin{tabular}{l|c|c|c|c|c|c}
\hline
Contributions & \textbf{Our work}\! & \cite{gao2014quantizer} & \cite{ciuonzo2020bandwidth} & \cite{wang2019detection} & \cite{li2019lr} & \cite{Mohammadi2022generalized}\\
\hline
\hline
\multicolumn{1}{m{0.3\linewidth}|}{Multi-bit quantization} & \checkmark & \checkmark & \checkmark &\checkmark & &\\
\hline
\multicolumn{1}{m{0.3\linewidth}|}{Raw quantization (RQ)}  & \checkmark & \checkmark & \checkmark &\checkmark & & \checkmark\\
\hline
\multicolumn{1}{m{0.3\linewidth}|}{LR quantization (LQ)} & \checkmark &  &  & & \checkmark & \checkmark\\
\hline
Sparsity model & \checkmark &  &  &\checkmark  & \checkmark & \checkmark\\
\hline
\multicolumn{1}{m{0.28\linewidth}|}{Error-prone reporting channel} & \checkmark & \checkmark & \checkmark &  & & \checkmark\\
\hline
\multicolumn{1}{m{0.3\linewidth}|}{Examine of the Fisher information loss} & \checkmark &  &  &  & & \\
\hline
\multicolumn{1}{m{0.28\linewidth}|}{Comparative  analysis between LQ and RQ} & \checkmark &  &  &  & & \\
\hline
\multicolumn{1}{m{0.28\linewidth}|}{Asymptotic relative efficiency analysis}& \checkmark &  &  & \checkmark & \checkmark &\\ 
\hline
\end{tabular}}
\end{table*}

The rest of the paper is structured as follows. Section \ref{Section2} derives the multi-bit distributed detection problems for sparse stochastic signals over non-ideal reporting channels, while Section \ref{Section3} details the  derivation of the proposed quantized detectors. In Section \ref{Section4}, we design the corresponding multi-bit quantization thresholds  and evaluate the performance degradation caused by quantization. Section \ref{Section5} presents the simulation results, followed by the concluding remarks in Section \ref{Section6}.

\textit{Notations:} Lowercase boldface letters are used to indicate vectors. $\mathbb{R}^{M \times N}$ denotes the space of  real matrices with dimension $M\times N$. Transposition is indicated by the superscript $(\cdot)^T$. $\mathbb{E}[\cdot]$ represents the statistical expectation. $I(\cdot)$ represents the indicator function. The notation $|\cdot|$ represents the absolute value of a real number. $|\!|\cdot|\!|$ denotes the Euclidean norm of a vector. The Dirac delta function is indicated by $\delta(\cdot)$. The symbols $\sim$ and $\stackrel a\sim$ stand for ``distributed as" and ``asymptotically distributed as", respectively. A Gaussian distribution with a mean of $\mu$ and a  variance of $\sigma^2$ is denoted by $\mathcal{N}(\mu,\sigma^2)$. The notation for probability mass functions is $P(\cdot)$, and its conditional counterpart is denoted as $P(\cdot|\cdot)$.
\section{System Model}\label{Section2}
\subsection{Signal Model}\label{Section2-1}
\begin{figure}[htb]
\centering
\vspace{-0.5cm}
\setlength{\abovecaptionskip}{0.cm}
\setlength{\abovecaptionskip}{0.cm}
\setlength{\belowdisplayskip}{3pt}
\includegraphics[width=\linewidth]{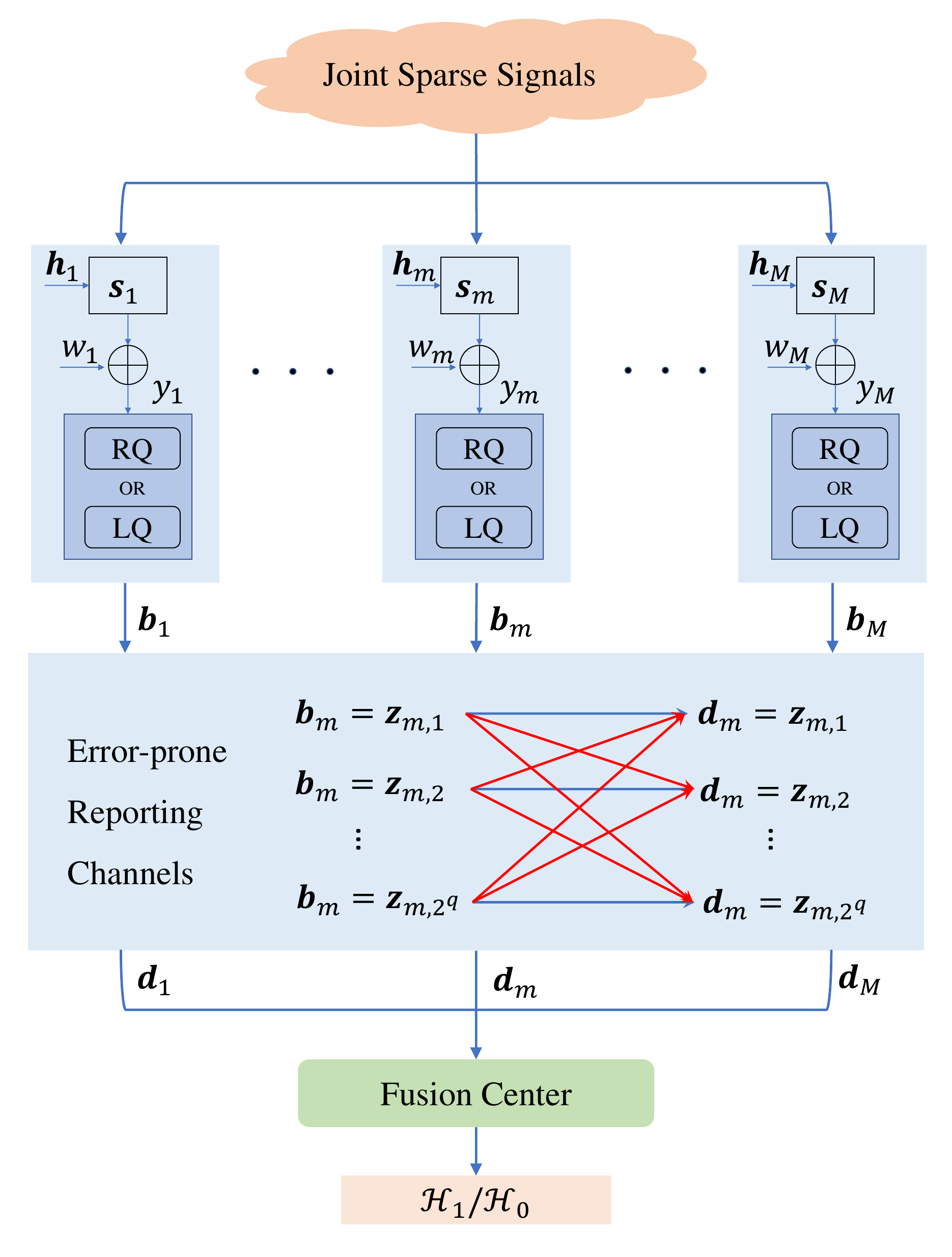}
\caption{Illustration of the distributed detection system for sparse signals with multi-bit quantization and error-prone reporting channels.}
\label{Figure1}
\end{figure}

Consider a distributed detection scheme comprising $M$ spatially located sensors and a FC. As observed from  Fig.\,\ref{Figure1},  each sensor simultaneously observes a phenomenon of interest that produces $N$-dimensional sparse signals, and linearly compresses the raw observation data using random measurement vectors. The task of detecting the presence of sparse signals based on compressed measurements can be formulated as the following binary hypothesis testing problem
\begin{align}\label{Eq1}
    \left\{
    \begin{array}{l}
      \mathcal{H}_0:y_m=w_m, \\
      \mathcal{H}_1:y_m=\pmb{h}_m^T\pmb{s}_m+w_m,
      \end{array}\right.
\end{align}
for $m=1,\cdots,M$, where $y_m\in\mathbb{R}$ denotes the compressed measurement of the $m$th sensor; $w_m$ represents the Gaussian noise with zero mean and a known variance $\sigma_w^2$; $\pmb{h}_m\in\mathbb{R}^{N\times1}$ is the signal-independent measurement vector; $\pmb{s}_m\in\mathbb{R}^{N\times1}$ denotes the sparse signal observed by the $m$th sensor. We assume that the sparse signals exhibit a joint sparsity structure, which can be modeled by the Bernoulli-Gaussian distribution.  Denote a $N\times1$ binary-valued vector, $\pmb{r}$, as the joint sparsity pattern with its entries assumed to be independent and identically distributed (i.i.d.) Bernoulli random variables which take the value 1 with probability $p$. The elemental structure of $\pmb{s}_m$ is consistent with that of $\pmb{r}$, and its non-zero entries follow a Gaussian distribution with a mean of zero and a variance of $\sigma_0^2$. The non-zero entries in $\pmb{s}_m$ are i.i.d. Bernoulli Gaussian variables adhering to the following probability distribution~\cite{Zayyani2016iterative,wang2018detection}
\begin{align}\label{Eq2}
   s_{m,n}\sim p\mathcal{N}(0,\sigma_0^2)+(1-p)\delta(s_{m,n}),\quad\forall m, \forall n.
\end{align}
with unknown sparsity degree $p$ and signal power $\sigma_s^2$. Consequently, the probability density function (PDF) of the measurement $y_m$ is distributed under $\mathcal{H}_0$ as
\begin{align}\label{Eq3}
	y_m|\mathcal{H}_0\sim \mathcal{N}(0,\sigma_w^2).
\end{align}
Based on the central limit theorem, the PDF of $y_m$ under $\mathcal{H}_1$ can be derived from \eqref{Eq1} and \eqref{Eq2} 
\begin{align}\label{Eq4}
	y_m|\mathcal{H}_1\stackrel a\sim\mathcal{N}(0,p\sigma_0^2|\!|\pmb{h}_m|\!|_2^2+\sigma_w^2).
\end{align} 
\subsection{Multi-bit Quantization at the Local Sensors}\label{Section2-2}
To conserve channel bandwidth and system power, a multi-bit quantizer is employed to quantize the measurement from the $m$-th sensor. Specifically, for a $q$-bit quantizer, the dynamic range of the sensor measurement is divided into $2^q$ non-overlapping quantization intervals, each being assigned with a binary codeword selected from the set $\{\pmb{z}_{m,1},\pmb{z}_{m,2},\cdots,\pmb{z}_{m,2^q}\}$, where $\pmb{z}_{m,i}=\left[z_{m,i,q},z_{m,i,q-1},\cdots,z_{m,i,1}\right]^T$ with $z_{m,i,k}\in\{0,1\}$. The endpoints of each interval are identified by the quantization thresholds  $\pmb{\tau}_m=\big[\tau_{m,1},\tau_{m,2},\cdots,\tau_{m,2^q-1}\big]^T$. By comparing the sensor output with the quantization thresholds, the quantizer determines the quantization interval and assigns the sensor measurement to the appropriate interval. After that, a binary codeword that matches the quantization interval is the output.
\subsection{Error-Prone Reporting Channels}\label{Section2-3}
The quantized data may encounter transmission errors when communicated to the FC. These errors are modeled using a $q$-ary discrete memoryless channel (DMC) model~\cite{djordjevic2019physical,daneshgaran2014ldpc} as shown in  Fig.\,\ref{Figure1}. We assume that the channels from different local sensors to the FC are independent, and the transmission of the $q$-bit quantized data sent by each sensor operates independently. Consequently, the transmission channel between sensors and the FC can be further modeled as $q$-ary  channels, where the transmission at each bit level is characterized as a BSC. Let $P_{\text{e},m}$ denote the crossover probability from the $m$-th sensor to the FC in the BSC. $P_{\text{e},m}$ can then be interpreted as the channel's average bit error rate (BER), which can be calculated using the classic Union bound algorithm \cite{sui2023space,sui2023performance}. The FC makes a global decision based on the received data regarding the presence or absence of the sparse signals. The problem at hand pertains to the design of the multi-bit quantizer in each sensor to assist the sparse signal detection at the FC.
\section{Two Proposed Quantized  LMPT Detectors}\label{Section3}
In this section, we design and construct two multi-bit LMPT detectors that fuse raw observations and local likelihood ratios, respectively. Then we analyze the asymptotic detection performance of the proposed quantized detectors.
\subsection{LMPT Based on Quantized Raw Observations}\label{Section3-1}
The compressed observation at each sensor node, i.e., $y_m$, is quantized into $q$ bits via raw quantization. Hence, the output codeword of the $q$-bit quantizer at the $m$th sensor can be expressed as
\begin{align}\label{Eq5}
	\pmb{b}_m^\text{rq}=
	\left\{
	\begin{array}{ll}
	\pmb{z}_{m,1}^\text{rq}\,, &-\infty<y_m<\tau_{m,1}^\text{rq},\\
	\pmb{z}_{m,2}^\text{rq}\,, &\tau_{m,1}^\text{rq}\leq y_m<\tau_{m,2}^\text{rq},\\
	\vdots                   &\vdots\\
	\pmb{z}_{m,2^q}^\text{rq}\,, &\tau_{m,2^q-1}^\text{rq}\leq y_m<+\infty.
	\end{array}\right.
\end{align}
Due to the distortion effect in BSC, the output $\pmb{b}_m^\text{rq}$ could be any  binary codeword. The probability of $\pmb{z}_{m,j}^\text{rq}$ being changed to $\pmb{z}_{m,i}^\text{rq}$ over the BSC can be calculated as
\begin{align}\label{Eq6}
	 P(\pmb{d}_{m}^\text{rq}=&\pmb{z}_{m,i}^\text{rq}|\pmb{b}_{m}^\text{rq}=\pmb{z}_{m,j}^\text{rq})\nonumber\\
	=&P_{\text{e},m}^{D_{m,i,j}^\text{rq}}(1-P_{\text{e},m})^{q-D_{m,i,j}^\text{rq}}\nonumber\\
	=&G(q,P_{\text{e},m},D_{m,i,j}^\text{rq}),
\end{align}
where $D_{m,i,j}^\text{rq}$ is the Hamming distance between $\pmb{z}_{m,j}^\text{rq}$ and $\pmb{z}_{m,i}^\text{rq}$, which can be defined as
\begin{align}\label{Eq7}
	D_{m,i,j}^\text{rq}\triangleq D(\pmb{z}_{m,i}^\text{rq},\pmb{z}_{m,j}^\text{rq})
	=q-\sum_{k=0}^{q-1}I(z_{m,i,k}^\text{rq},z_{m,j,k}^\text{rq}).
\end{align}

The probability mass function (PMF) of the output $\pmb{d}_m^\text{rq}$ with the BSC under $\mathcal{H}_1$ is given by
\begin{align}\label{Eq8}
	P(\pmb{d}_m^\text{rq};p)&=\sum_{i=1}^{2^q}\sum_{j=1}^{2^q}P(\pmb{d}_m^\text{rq}=\pmb{z}_{m,i}^\text{rq}|\pmb{b}_m^\text{rq}=\pmb{z}_{m,j}^\text{rq})P(\pmb{b}_m^{\text{rq}}=\pmb{z}_{m,j}^\text{rq})\nonumber\\
	&=\prod_{i=1}^{2^q}\left[\sum_{j=1}^{2^q}G(q,P_{\text{e},m},D_{m,i,j}^\text{rq})Q_{m,j}^\text{rq}(p)\right]^{I(\pmb{d}_m^\text{rq}=\pmb{z}_{m,i}^\text{rq})},
\end{align}
where
\begin{align}\label{Eq9}
Q_{m,j}^\text{rq}(p)&\triangleq P(\pmb{b}_m^{\text{rq}}=\pmb{z}_{m,j}^\text{rq};p)
=P(\tau_{m,j-1}^\text{rq}\leq y_m<\tau_{m,j}^\text{rq})\nonumber\\
&=\Phi\left(\frac{\tau_{m,j-1}^\text{rq}}{\sigma_m(p,\sigma_0^2)}\right)-\Phi\left(\frac{\tau_{m,j}^\text{rq}}{\sigma_m(p,\sigma_0^2)}\right),
\end{align}
with $\Phi(\beta)=1/\sqrt{2\pi}\int_{\beta}^{+\infty}\exp(-\alpha^2/2)\,\text{d}\alpha$ being the complementary cumulative density function (CCDF) of $w_{m}$, and $\sigma_m^2(p,\sigma_0^2)\triangleq p\sigma_0^2|\!|\pmb{h}_m|\!|_2^2+\sigma_w^2$.
Denote the data collected by the FC as  $\pmb{d}^\text{rq}=\left[ \pmb{d}_1^\text{rq},\pmb{d}_2^\text{rq},\cdots,\pmb{d}_M^\text{rq}\right]$. The likelihood function under $\mathcal{H}_1$ can be expressed as
\begin{multline}\label{Eq10}
\!\!P(\pmb{d}^\text{rq}|\mathcal{H}_1;p)\\
\!=\prod_{m=1}^{M}\!\prod_{i=1}^{2^q}\left[\sum_{j=1}^{2^q}\!G(q,\!P_{\text{e},m},D_{m,i,j}^\text{rq})Q_{m,j}^\text{rq}(p)\right]^{\!I(\pmb{d}_m^\text{rq},\pmb{z}_{m,i}^\text{rq})}.
\end{multline}
Based on the quantized raw observations, the LMPT detector \cite{kay1998fundamentals} can be obtained as 
\begin{equation}\label{Eq11}
T_\text{LMPT}^\text{rq}=\left(\frac{\partial\ln P(\pmb{d}^\text{rq}|\mathcal{H}_1;p)}{\partial p}\bigg/\sqrt{\text{FI}_q^\text{rq}(p)}\right)_{p=0}\underset{H_0}{\mathop{\overset{H_1}{\mathop{\gtrless}}\,}}\eta^\text{rq},
\end{equation}
where {$\eta^\text{rq}$ is the detection threshold.} Taking the derivative of the logarithm of \eqref{Eq10} with respect to $p$ leads to (see Appendix \ref{Appendices1} for details)
\begin{multline}\label{Eq12}
\!\frac{\partial\ln P(\pmb{d}^\text{rq}|\mathcal{H}_1;p)}{\partial p}=\frac{\sigma_0^2}{2}\sum_{m=1}^{M}\frac{|\!|\pmb{h}_m|\!|_2^2}{\sigma_m^3(p,\sigma_0^2)}\\
\!\times\sum_{i=1}^{2^q}\left[\frac{\!I(\pmb{d}_m^\text{rq},\pmb{z}_{m,i}^\text{rq})\!\sum_{j=1}^{2^q}\!G(q,P_{\text{e},m},D_{m,i,j}^\text{rq})F_{m,j}^\text{rq}(p)}{\sum_{j=1}^{2^q}\!G(q,P_{\text{e},m},D_{m,i,j}^\text{rq})Q_{m,j}^\text{rq}(p)}\right],
\end{multline}
where 
\begin{align}\label{Eq13}
F_{m,j}^\text{rq}(p)&=\tau_{m,j-1}^\text{rq}\Psi\big({\tau_{m,j-1}^\text{rq}}/\textstyle{\sigma_m(p,\sigma_0^2)}\big)\nonumber\\
&-\tau_{m,j}^\text{rq}\Psi\big({\tau_{m,j}^\text{rq}}/\textstyle{\sigma_m(p,\sigma_0^2)}\big).
\end{align}
In \eqref{Eq11}, $\text{FI}_q^\text{rq}(p)$ denotes the Fisher information, which can be formulated as (see Appendix \ref{Appendices2} for details)
\begin{multline}\label{Eq14}
\text{FI}_q^\text{rq}(p)
\triangleq\!-\mathbb{E}\left[\!\frac{\partial^2\ln P(\pmb{d}^\text{rq}|\mathcal{H}_1;p)}{\partial p^2}\right]
=\frac{\sigma_0^4}{4}\sum_{m=1}^{M}\frac{|\!|\pmb{h}_m|\!|_2^4}{\sigma_m^6(p,\sigma_0^2)}\\
\times\sum_{i=1}^{2^q}\frac{\left[\sum_{j=1}^{2^q}G(q,P_{\text{e},m},D_{m,i,j}^\text{rq})F_{m,j}^\text{rq}(p)\right]^2}{\sum_{j=1}^{2^q}G(q,P_{\text{e},m},D_{m,i,j}^\text{rq})Q_{m,j}^\text{rq}(p)}.
\end{multline}
Note that although the FI is a scalar that is independent of measurements and, hence, can be dropped, it is retained in \eqref{Eq11} as the scaled test variable has a simple asymptotic distribution as shown in Section \ref{Section3-3}. 

By substituting $p=0$ into \eqref{Eq12} and \eqref{Eq14}, the LMPT detector based on quantized raw observations is given by 
\begin{multline}\label{Eq15}
	T_\text{LMPT}^\text{rq}\propto\sum_{m=1}^{M}\frac{|\!|\pmb{h}_m|\!|_2^2}{\sigma_w^3}\sum_{i=1}^{2^q}I(\pmb{d}_m^\text{rq},\pmb{z}_{m,i}^\text{rq})\\
	\times\frac{\sum_{j=1}^{2^q}G(q,P_{\text{e},m},D_{m,i,j}^\text{rq})F_{m,j}^\text{rq}(0)}{\sum_{j=1}^{2^q}G(q,P_{\text{e},m},D_{m,i,j}^\text{rq})Q_{m,j}^\text{rq}(0)},
\end{multline}
where we have
\begin{align}\label{Eq16}
\!F_{m,j}^\text{rq}(0)=\tau_{m,j-1}^\text{rq}\Psi\left({\tau_{m,j-1}^\text{rq}}/{\sigma_w}\right)-\tau_{m,j}^\text{rq}\Psi\left({\tau_{m,j}^\text{rq}}/{\sigma_w}\right),\!
\end{align}
\begin{align}\label{Eq17}
Q_{m,j}^\text{rq}(0)=\Phi\left({\tau_{m,j-1}^\text{rq}}/{\sigma_w}\right)-\Phi\left({\tau_{m,j}^\text{rq}}/{\sigma_w}\right).
\end{align}
\subsection{LMPT Based on Quantized Likelihood Ratios}\label{Section3-2}
In this subsection, we develop the LMPT detector, which utilizes $q$-bit quantizer to fuse local LRs instead of  using the quantized raw observations directly. Specifically, the local LR can be obtained from \eqref{Eq4} as follows:
\begin{align}\label{Eq18}
\mathcal{L}_m(y_m;p)&=P(y_m|\mathcal{H}_1;p)/P(y_m|\mathcal{H}_0)\nonumber\\
&=\sqrt{\frac{\sigma_w^2}{\sigma_m^2(p,\sigma_0^2)}}\exp\left(\frac{p\sigma_0^2|\!|\pmb{h}_m|\!|_2^2}{2\sigma_w^2\sigma_m^2(p,\sigma_0^2)}y_m^2\right).
\end{align}
Based on \eqref{Eq18}, $\mathcal{L}_m(y_m;p)$ is a monotonically increasing function of $y_m^2$ or $|y_m|$. Consequently, at the $m$th sensor, the $q$-bit quantizer based on the LR can be expressed as
\begin{align}\label{Eq19}
\pmb{b}_m^\text{lq}=
\left\{
\begin{array}{ll}
\pmb{z}_{m,1}^\text{lq}\,, &0<|y_m|<\tau_{m,1}^\text{lq},\\
\pmb{z}_{m,2}^\text{lq}\,, &\tau_{m,1}^\text{lq}\leq |y_m|<\tau_{m,2}^\text{lq},\\
\vdots                   &\vdots\\
\pmb{z}_{m,2^q}^\text{lq}\,, &\tau_{m,2^q-1}^\text{lq}\leq |y_m|<+\infty.
\end{array}\right.
\end{align}
Similarly, the quantized data is sent to the FC via BSC, and the probability that $\pmb{z}_{m,j}^\text{lq}$ is changed to $\pmb{z}_{m,i}^\text{lq}$ over the BSC can be computed as
\begin{align}\label{Eq20}
P(\pmb{d}_{m}^\text{lq}=&\pmb{z}_{m,i}^\text{lq}|\pmb{b}_{m}^\text{lq}=\pmb{z}_{m,j}^\text{lq})\nonumber\\
=&P_{\text{e},m}^{D_{m,i,j}^\text{lq}}(1-P_{\text{e},m})^{q-D_{m,i,j}^\text{lq}}\nonumber\\
\triangleq&G(q,P_{\text{e},m},D_{m,i,j}^\text{lq}),
\end{align}
where $D_{m,i,j}^\text{lq}$ denotes the Hamming distance between $\pmb{z}_{m,j}^\text{lq}$ and $\pmb{z}_{m,i}^\text{lq}$, and is defined similarly to that of \eqref{Eq7}. The PMF of the received data $\pmb{d}_m^\text{lq}$ can be derived as
\begin{align}\label{Eq21}
\!P(\pmb{d}_m^\text{lq};p)\!&=\sum_{i=1}^{2^q}\sum_{j=1}^{2^q}P(\pmb{d}_m^\text{lq}=\pmb{z}_{m,i}^\text{lq}|\pmb{b}_m^\text{lq}=\pmb{z}_{m,j}^\text{lq})P(\pmb{b}_m^{\text{lq}}=\pmb{z}_{m,j}^\text{lq})\nonumber\\
&=\prod_{i=1}^{2^q}\left[2\sum_{j=1}^{2^q}G(q,\!P_{\text{e},m},D_{m,i,j}^\text{lq})Q_{m,j}^\text{lq}(p)\!\right]^{\!I(\pmb{d}_m^\text{lq}=\pmb{z}_{m,i}^\text{lq})},
\end{align}
where
\begin{align}\label{Eq22}
Q_{m,j}^\text{lq}(p)&\triangleq \frac{1}{2}P(\pmb{b}_m^\text{lq}=\pmb{z}_{m,j}^\text{lq};p)
=\frac{1}{2}P(\tau_{m,j-1}^\text{lq}\leq |y_m|<\tau_{m,j}^\text{lq})\nonumber\\
&=\Phi\left(\frac{\tau_{m,j-1}^\text{lq}}{\sigma_m(p,\sigma_0^2)}\right)-\Phi\left(\frac{\tau_{m,j}^\text{lq}}{\sigma_m(p,\sigma_0^2)}\right).
\end{align}
The likelihood function of the data received by the FC can be expressed as
\begin{multline}\label{Eq23}
\!P(\pmb{d}^\text{lq}|\mathcal{H}_1;p)\\
\!=\prod_{m=1}^{M}\!\prod_{i=1}^{2^q}\left[2\sum_{j=1}^{2^q}\!G(q,\!P_{\text{e},m},D_{m,i,j}^\text{lq})Q_{m,j}^\text{lq}(p)\!\right]^{\!I(\pmb{d}_m^\text{lq},\pmb{z}_{m,i}^\text{lq})\!}.
\end{multline}
Then we take the derivative of the logarithm of \eqref{Eq23} with respect to $p$, yielding (see Appendix \ref{Appendices3} for details)
\begin{multline}\label{Eq24}
\frac{\partial\ln P(\pmb{d}^\text{lq}|\mathcal{H}_1;p)}{\partial p}=\frac{\sigma_0^2}{2}\sum_{m=1}^{M}\frac{|\!|\pmb{h}_m|\!|_2^2}{\sigma_m^3(p,\sigma_0^2)}\\
\times\sum_{i=1}^{2^q}\left[\!\frac{\!I(\pmb{d}_m^\text{lq},\pmb{z}_{m,i}^\text{lq})\sum_{j=1}^{2^q}G(q,P_{\text{e},m},D_{m,i,j}^\text{lq})F_{m,j}^\text{lq}(p)}{\sum_{j=1}^{2^q}G(q,P_{\text{e},m},D_{m,i,j}^\text{lq})Q_{m,j}^\text{lq}(p)}\right],
\end{multline}
where 
\begin{align}\label{Eq25}
F_{m,j}^\text{lq}(p)&=\tau_{m,j-1}^\text{lq}\Psi\big({\tau_{m,j-1}^\text{lq}}/\textstyle{\sigma_m(p,\sigma_0^2)}\big)\nonumber\\
&-\tau_{m,j}^\text{lq}\Psi\big({\tau_{m,j}^\text{lq}}/\textstyle{\sigma_m(p,\sigma_0^2)}\big).
\end{align}
The Fisher information is formulated as (see Appendix \ref{Appendices4} for details)
\begin{multline}\label{Eq26}
\text{FI}_q^\text{lq}(p)
\triangleq\!\mathbb{E}\left[\!\left(\frac{\partial\ln P(\pmb{d}^\text{lq}|\mathcal{H}_1;p)}{\partial p}\right)^2\right]
=\frac{\sigma_0^4}{2}\sum_{m=1}^{M}\frac{|\!|\pmb{h}_m|\!|_2^4}{\sigma_m^6(p,\sigma_0^2)}\\
\times\sum_{i=1}^{2^q}\frac{\left[\sum_{j=1}^{2^q}G(q,P_{\text{e},m},D_{m,i,j}^\text{lq})F_{m,j}^\text{lq}(p)\right]^2}{\sum_{j=1}^{2^q}G(q,P_{\text{e},m},D_{m,i,j}^\text{lq})Q_{m,j}^\text{lq}(p)}.
\end{multline}
Upon substituting $p=0$ into \eqref{Eq24} and \eqref{Eq26}, the LMPT detector based on quantized raw observations is given by 
\begin{multline}\label{Eq27}
T_\text{LMPT}^\text{lq}\propto\sum_{m=1}^{M}\frac{|\!|\pmb{h}_m|\!|_2^2}{\sigma_w^3}\sum_{i=1}^{2^q}I(\pmb{d}_m^\text{lq},\pmb{z}_{m,i}^\text{lq})\\
\times\frac{\sum_{j=1}^{2^q}G(q,P_{\text{e},m},D_{m,i,j}^\text{lq})F_{m,j}^\text{lq}(0)}{\sum_{j=1}^{2^q}G(q,P_{\text{e},m},D_{m,i,j}^\text{lq})Q_{m,j}^\text{lq}(0)},
\end{multline}
where
\begin{align}\label{Eq28}
\!F_{m,j}^\text{lq}(0)=\tau_{m,j-1}^\text{lq}\Psi\big({\tau_{m,j-1}^\text{lq}}/{\sigma_w}\big)-\tau_{m,j}^\text{lq}\Psi\big({\tau_{m,j}^\text{lq}}/{\sigma_w}\big),
\end{align}
\begin{align}\label{Eq29}
Q_{m,j}^\text{lq}(0)=\Phi\big({\tau_{m,j-1}^\text{lq}}/{\sigma_w}\big)-\Phi\big({\tau_{m,j}^\text{lq}}/{\sigma_w}\big).
\end{align}
 \subsection{Asymptotic Detection Performance}\label{Section3-3}
According to \cite{kay1998fundamentals}, the asymptotic distribution of the LMPT test statics $T_\text{LMPT}$ in \eqref{Eq15} and \eqref{Eq27} can be obtained as
\begin{align}\label{Eq30}
T_\text{LMPT}\stackrel a\sim\left\{
\begin{array}{ll}
\mathcal{N}(0,1), &\text{under}\quad \mathcal{H}_0\\
\mathcal{N}(\lambda_q,1), &\text{under}\quad \mathcal{H}_1
\end{array}\right.,
\end{align}
where 
\begin{align}\label{Eq31}
\lambda_q=p\sqrt{\text{FI}_q(0)}
\end{align}
denotes the non-centrality parameter. For the sake of simplicity, the superscripts ``rq" and ``lq" are omitted whenever there is no confusion. It is readily observed from \eqref{Eq30} that for a given threshold $\eta$, the probability of false alarm can be expressed as
\begin{align}\label{Eq32}
	P_\text{FA}=P(T_\text{LMPT}>\eta|\mathcal{H}_0)=\Phi(\eta).
\end{align} 
Similarly, the probability of detection can be written as
\begin{align}\label{Eq33}
	P_\text{D}=P(T_\text{LMPT}>\eta|\mathcal{H}_1)=\Phi_{\lambda_q}(\eta),
\end{align}
where $\Phi_{\lambda_q}(\beta)=1/\sqrt{2\pi}\int_{\beta}^{+\infty}\exp(-(\alpha-\lambda_q)^2/2)\,{\text{d}}\alpha$ denotes the CCDF for a non-central normal distribution with  non-centrality parameter $\lambda_q$. 

\textit{Remarks:} Referring to \eqref{Eq31}, we can obtain the detection threshold $\eta$ given a specific probability of false alarm $P_\text{FA}$. Then, the probability of detection $P_\text{D}$ can be calculated by substituting $\eta$ into \eqref{Eq33}.  As demonstrated in \eqref{Eq32} and \eqref{Eq33}, the detection performance improves with higher $\lambda_q$, while \eqref{Eq31} suggests that $\lambda_q$  increases monotonically with $\text{FI}_q$, which is a function of the quantization threshold and the crossover probability. Therefore, by maximizing $\text{FI}_q$, we can determine a set of quantization thresholds that allow the quantized LMPT detectors to achieve near-optimal detection performance for a given $P_{\text{e},m}$.
\section{Design of Quantization Threshold and Performance Analysis}\label{Section4}
In this section, we first design quantization thresholds for the proposed quantized LMPT detectors to achieve near-optimal asymptotic performance. The normalized Fisher information and asymptotic relative efficiency are then derived analytically to analyze and compensate for the loss of information resulting from quantization. A comparative analysis between RQ and LQ is also provided to account for their differences.
\subsection{Multi-bit Quantization Threshold Design}\label{Section4-1}
Inspired by the analysis in Section \ref{Section3-3}, the quantization threshold design can be turned into an optimization problem  with regard to FI as shown below
\begin{align}\label{Eq34}
	\!\underset{\{\!\pmb{\tau}_m\!\}_{m=\!1}^{M}}{\!\max}\!\sum_{m=1}^{M}\!\frac{|\!|\pmb{h}_m|\!|_2^4}{\sigma_w^6}\!
	\sum_{i=1}^{2^q}\frac{\big[\sum_{j=1}^{2^q}\!G(q,\!P_{\text{e},m},\!D_{m,i,j})F_{m,j}(0)\big]^2}{\sum_{j=1}^{2^q}\!G(q,\!P_{\text{e},m},\!D_{m,i,j})Q_{m,j}(0)}.\!
\end{align}
Assuming that the reporting channels between each local sensor and the FC are independent, the optimization issue indicated in \eqref{Eq34} can be divided into $M$ separate  problems, yielding
\begin{align}\label{Eq35}
\underset{\{\!\pmb{\tau}_m\!\}_{m=\!1}^{M}}{\!\max}\sum_{i=1}^{2^q}\frac{\big[\sum_{j=1}^{2^q}\!G(q,\!P_{\text{e},m},\!D_{m,i,j})F_{m,j}(0)\big]^2}{\sum_{j=1}^{2^q}\!G(q,\!P_{\text{e},m},\!D_{m,i,j})Q_{m,j}(0)}\!\nonumber\\
\text{s.t.\quad} \tau_{m,0}<\tau_{m,1}<\cdots<\tau_{m,2^q-1}<+\infty,
\end{align}
where $\tau_{m,0}^\text{rq}=-\infty$ and $\tau_{m,0}^\text{lq}=0$. The objective function of the optimization problem described by \eqref{Eq35} is a nonlinear non-convex function whose closed form solution is difficult to obtain. While a gradient search method can be used to solve the problem, it still relies on the selection of stochastic initial points and has a high complexity. As a result, a particle swarm optimization approach with low computing cost is used in this research to solve \eqref{Eq35}.

Complexity requirements: similar to that of~\cite{ciuonzo2020bandwidth}, it is assumed that all pre-computations not dependent on $\pmb{y} $ have been previously carried out and stored in memory at the FC. The overall  complexity of the optimization problem in \eqref{Eq35} is primarily determined by the number of iterations and the complexity in the numerator and denominator. Given a specific value of $m$, the complexity for a singular computation of the objective function as depicted in (35) is $\mathcal{O}(4^q)$. 
The complexity associated with solving the optimization problem via PSOA is  primarily contingent on the fitness evaluation during the initialization of the swarm and throughout the iterative procedure~\cite{wang2018particle}. Let $Q$ represent the number of particles generated during the initialization phase, and $T$ denote the total number of iterations. Then the complexity for addressing \eqref{Eq35} using PSOA is roughly $\mathcal{O}(Q4^q) + \mathcal{O}(TQ4^q)$. 
\subsection{Derivation of the Clairvoyant LMPT Detector}\label{Section4-2}
Based on raw observations without quantization, the Clairvoyant LMPT detector can be expressed as
\begin{align}\label{Eq36}
	\tilde{T}(\pmb{y})=\left(\frac{\partial\ln {P}(\pmb{y}|\mathcal{H}_1;p)}{\partial p}\bigg/\sqrt{\text{FI}(p)}\right)_{p=0}\underset{H_0}{\mathop{\overset{H_1}{\mathop{\gtrless}}\,}}\tilde{\eta},
\end{align}
where $\pmb{y}\triangleq\left[y_1,y_2,\cdots,y_M\right]^T$ denotes the raw observations received by the FC via a perfect channel, $\text{FI}(p)$ denotes the Fisher information, and $\tilde{\eta}$ is the detection threshold. According to \eqref{Eq1}, the likelihood function under $\mathcal{H}_1$ is given by
\begin{align}\label{Eq37}
	\!\!P(\pmb{y}|\mathcal{H}_1;p)\!=\!\prod_{m=1}^{M}\!\frac{1}{\sqrt{2\pi\sigma_m^2(p,\sigma_0^2)}}\exp\left\lbrace \!-\frac{y_m^2}{2\sigma_m^2(p,\sigma_0^2)}\!\right\rbrace.
\end{align}
Then we can obtain the derivative of the logarithm of \eqref{Eq37} with respect to $p$, yielding
\begin{align}\label{Eq38}
\frac{\partial\ln P(\pmb{y}|\mathcal{H}_1;p)}{\partial p}=\frac{\sigma_0^2}{2}\sum_{m=1}^{M}\frac{|\!|\pmb{h}_m|\!|_2^2\left[y_m^2-\sigma_m^2(p,\sigma_0^2)\right]}{\sigma_m^4(p,\sigma_0^2)}.
\end{align}
The Fisher information is calculated as
\begin{align}\label{Eq39}
\text{FI}(p)\!
\triangleq\!\mathbb{E}\left[\!\left(\frac{\partial\ln\!P(\pmb{y}|\mathcal{H}_1;p)}{\partial p}\right)^2\right]
\!=\!\frac{\sigma_0^4}{2}\sum_{m=1}^{M}\frac{|\!|\pmb{h}_m|\!|_2^4}{\sigma_m^4(p,\sigma_0^2)}.	
\end{align}
Substituting $p=0$ into \eqref{Eq38} and \eqref{Eq39}, the LMPT detector based on unquantized raw observations is given by 
\begin{align}\label{Eq40}
\tilde{T}(\pmb{y})\propto\sum_{m=1}^{M}\frac{|\!|\pmb{h}_m|\!|_2^2(y_m^2-\sigma_w^2)}{\sqrt{\sum_{m=1}^{M}|\!|\pmb{h}_m|\!|_2^4}}.
\end{align}
Similarly, the asymptotic distribution of the LMPT test statistics $\tilde{T}(\pmb{y})$ in \eqref{Eq40} can be obtained as follows
\begin{align}\label{Eq41}
\tilde{T}(\pmb{y})\stackrel a\sim\left\{
\begin{array}{ll}
\mathcal{N}(0,1), &\text{under}\quad \mathcal{H}_0\\
\mathcal{N}(\tilde{\lambda},1), &\text{under}\quad \mathcal{H}_1
\end{array}\right.,
\end{align}
where 
\begin{align}\label{Eq42}
\tilde{\lambda}=p\sqrt{\text{FI}(0)}
\end{align}
denotes the non-centrality parameter. 
\subsection{Analysis of the Fisher Information Loss}\label{Section4-3}
We then examine the Fisher information loss caused by quantization. By assuming a homogeneous environment, the sensors are supposed to share the same observation qualities with $|\!|\pmb{h}_m|\!|_2^2=|\!|\pmb{h}|\!|_2^2$ and $P_{\text{e},m}=P_{\text{e}}, \forall m$. In this context, we can observe from \eqref{Eq14} and \eqref{Eq39} that the normalized Fisher information
\begin{align}\label{Eq43}
	\frac{\text{FI}_q^\text{rq}(0)}{\text{FI}(0)}=\frac{1}{2}\sum_{i=1}^{2^q}\frac{\left[\sum_{j=1}^{2^q}G(q,P_{\text{e},m},D_{m,i,j}^\text{rq})\bar{F}_{m,j}^\text{rq}(0)\right]^2}{\sum_{j=1}^{2^q}G(q,P_{\text{e},m},D_{m,i,j}^\text{rq})Q_{m,j}^\text{rq}(0)},
\end{align}
where $\bar{F}_{m,j}^\text{rq}(0)=F_{m,j}^\text{rq}(0)/\sigma_w$. Similarly, from \eqref{Eq26} and \eqref{Eq39}, we can conclude that
\begin{align}\label{Eq44}
	\frac{\text{FI}_q^\text{lq}(0)}{\text{FI}(0)}=\sum_{i=1}^{2^q}\frac{\left[\sum_{j=1}^{2^q}G(q,P_{\text{e},m},D_{m,i,j}^\text{lq})\bar{F}_{m,j}^\text{lq}(0)\right]^2}{\sum_{j=1}^{2^q}G(q,P_{\text{e},m},D_{m,i,j}^\text{lq})Q_{m,j}^\text{lq}(0)}.
\end{align}

\begin{figure}[htbp]
\centering
\includegraphics[width=\linewidth]{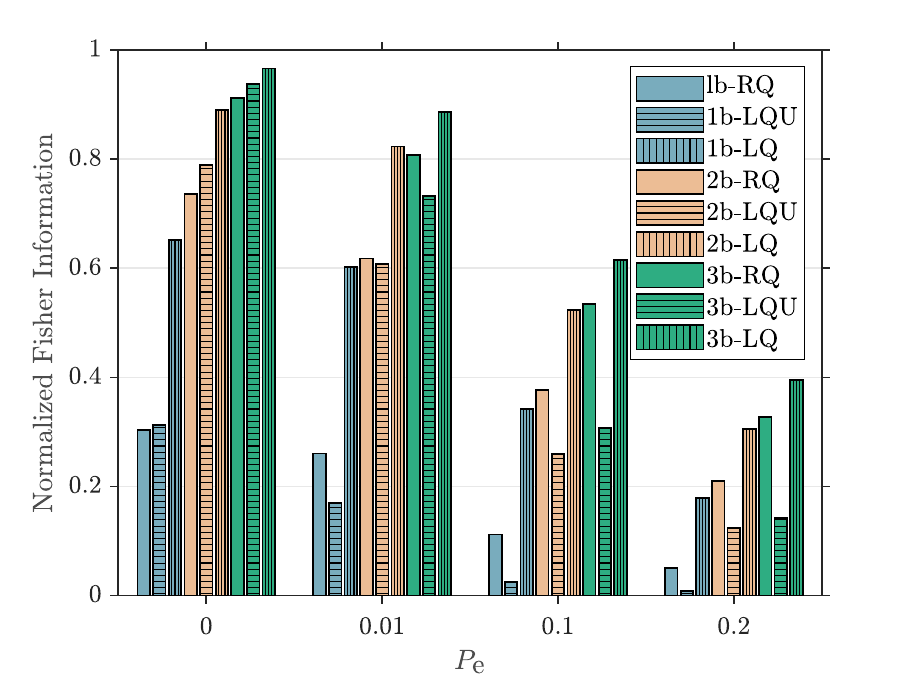}
\caption{Normalized Fisher information of RQ, LQU, and LQ with different values of $q$ and $P_\text{e}$.}
\label{Figure2}
\end{figure}

Figure\,\ref{Figure2} illustrates the Fisher information under various channel conditions. In the figure, ``1b'', ``2b'' and ``3b'' denote the quantization depth, while ``RQ'', ``LQU'' and ``LQ'' represent the  raw quantizer with optimized thresholds, LR quantizer with uniform thresholds, and the LR quantizers with optimized thresholds, respectively. As shown in Fig.\,\ref{Figure2}, when employing the optimized threshold, LQ is capable of striking highest normalized FI in all scenarios, while the normalized FI of 1b-RQ is approximately half of 1b-LQ. This can be readily verified by referring to \eqref{Eq43} and \eqref{Eq44}, as their quantization thresholds are close to one another at this point. Additionally, it was found that the normalized  FI of LQU declines dramatically as the cross probability $P_\text{e}$ increases. This suggests that a set threshold, which does not adjust in response to the channel state, cannot provide sufficient information for FC to make decisions. 
\begin{figure}[htbp]
\centering
\includegraphics[width=\linewidth]{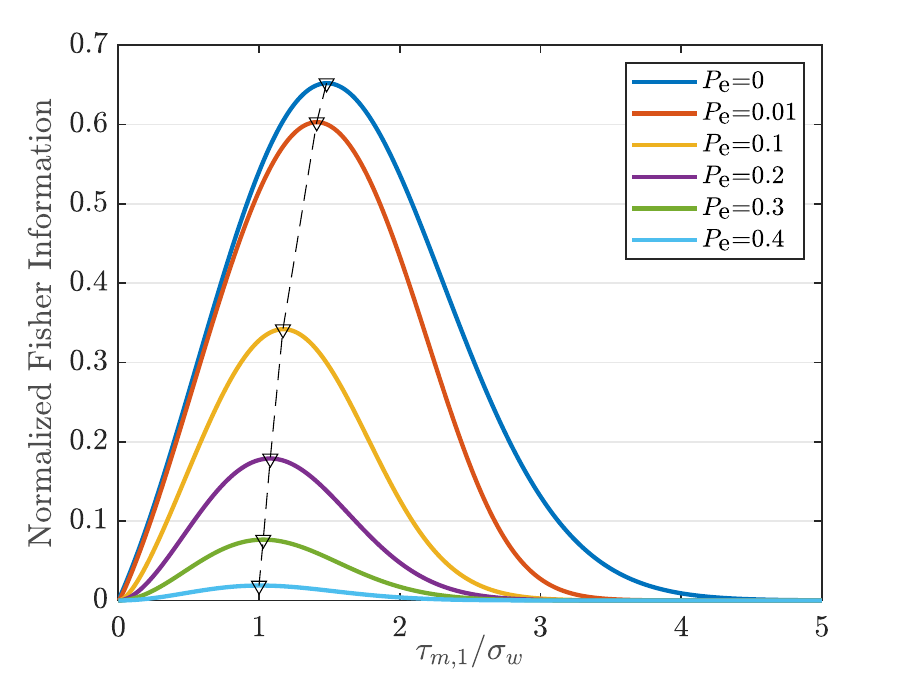}
\caption{Normalized Fisher information of 1b-LQ versus normalized threshold $\tau_{m,1}/\sigma_w$ with different values of $P_\text{e}$.}
\label{Figure3}
\end{figure}

In Fig.\,\ref{Figure3}, we plot the normalized Fisher information of 1b-LQ versus normalized threshold $\bar{\tau}_{m,1}\triangleq{\tau_{m,1}}/{\sigma_w}$ with different values of $P_\text{e}$. It is seen that the optimal threshold that corresponds to the maximum normalized Fisher information decreases as the increase of $P_\text{e}$ (when $P_\text{e}<0.5$), indicating that for 1b-LQ each sensor must lower its quantization threshold under more challenging channel circumstances in order to provide FC with more information. This also gives us a means to find quantization thresholds for 1b-LQ when solving the optimization problem in \eqref{Eq35} is not practical.

\subsection{Asymptotic Relative Efficiency}\label{Section4-4}
In order to compensate for the information loss caused  by quantization, the quantized detectors require more sensors to attain the same detection performance as the Clairvoyant detector. The ratio between the number of sensors needed to attain the same performance for a quantized detector and a clairvoyant detector is known as the ARE. Let $\tilde{M}$ and $M$ be the number of sensors used in the $\tilde{T}(\pmb{y})$ and $T_\text{LMPT}$ tests, respectively. To ensure that both have the same detection performance, it is required that
\begin{align}\label{Eq45}
	\tilde{\lambda}=\lambda_q.
\end{align}
By  substituting \eqref{Eq16} for RQ or \eqref{Eq28} for LQ into \eqref{Eq31}  and combining \eqref{Eq42} and \eqref{Eq45}, it is clear that the ARE of the two in a homogeneous environment can be expressed as
\begin{align}\label{Eq46}
	r^{\text{rq}}
	=\!2\left\lbrace\sum_{i=1}^{2^q}\frac{\left[\sum_{j=1}^{2^q}G(q,P_\text{e},D_{m,i,j}^\text{rq})\bar{F}_{m,j}^\text{rq}(0)\right]^2}{\sum_{j=1}^{2^q}G(q,P_\text{e},D_{m,i,j}^\text{rq})Q_{m,j}^\text{rq}(0)}\!\right\rbrace^{-1},\!
\end{align}
\begin{align}\label{Eq47}
	r^{\text{lq}}=\left\lbrace\sum_{i=1}^{2^q}\frac{\left[\sum_{j=1}^{2^q}G(q,P_\text{e},D_{m,i,j}^\text{lq})\bar{F}_{m,j}^\text{lq}(0)\right]^2}{\sum_{j=1}^{2^q}G(q,P_\text{e},D_{m,i,j}^\text{lq})Q_{m,j}^\text{lq}(0)}\right\rbrace^{-1},
\end{align}
where $r^\text{rq}$ and $r^\text{lq}$ stand for the ARE of RQ and LQ, respectively. Table~\ref{table2} displays the ARE for RQ, LQU, and LQ under various $q$ and $P_\text{e}$ conditions. From Table~\ref{table2} we can conclude that the number of sensors needed to attain the same detection performance gradually drops as $q$ grows. It is also observed that LQ requires the fewest sensors for the same $q$ and $P_\text{e}$ compared to other counterparts.
\begin{table*}[t]
\renewcommand{\arraystretch}{1.25}
\footnotesize
\centering
\caption{The ARE For RQ, LQ, and LQU under various $q$ and $P_{\text{e}}$ conditions.}
\label{table2}
\begin{tabular}{l|c|c|c|c|c|c|c|c|c|c|c|c}
\hline
$q$ & \multicolumn{4}{c|}{ 1} & \multicolumn{4}{c|}{ 2}  &\multicolumn{4}{c}{3} \\
\hline
\hline
$P_\text{e}$ & 0 & 0.01 & 0.1 & 0.2  & 0 & 0.01 & 0.1 & 0.2  & 0 & 0.01 & 0.1 & 0.2 \\
\hline
$r^\text{rq}$ & 3.29 &	3.84	&8.91	&19.53&	1.36&	1.62	&2.65	&4.76&	1.10&	1.24	&1.87	&3.06\\
\hline
$r^\text{lqu}$      & 3.19	&5.88	&39.84	&119.05	&1.35	&1.64	&3.85	&8.09	&1.07&	1.37&	3.26&	7.01 \\
\hline
$r^\text{lq}$ & 1.53 &	1.66&	2.92&	5.59	&1.12	&1.21	&1.91	&3.28	&1.03&	1.13	&1.63	&2.53\\
\hline
\end{tabular}
\end{table*}
\subsection{Comparison of RQ and LQ}\label{Section4-5}
The primary distinction between the raw quantizer in \eqref{Eq5} and the LR quantizer in \eqref{Eq19} is that the latter quantizes the absolute value of the observed measurement. The above distinction results in differences in the division of the quantization interval and the encoding of the intervals into binary code words. Specifically, LQ divides intervals more finely, resulting in a larger Hamming distance between the binary codewords corresponding to the low and high-probability regions. These characteristics make it less likely for LQ to transmit information erroneously under the same $P_\text{e}$ in error-prone channels. Taking $q=2$ as an example, the aforementioned variances will be further elucidated.
\begin{figure}[htb]
\centering
\includegraphics[width=\linewidth]{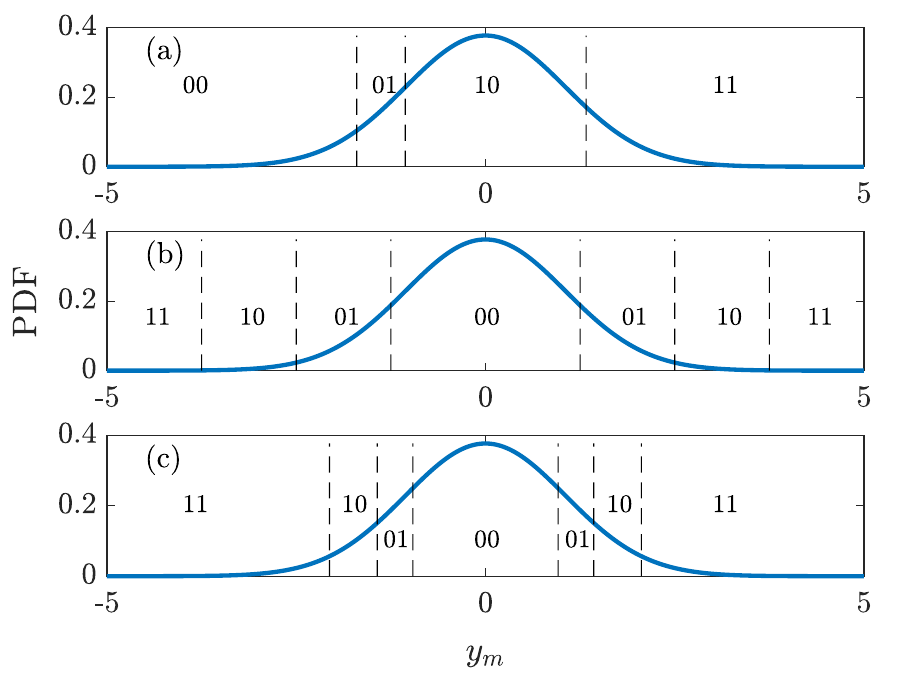}
\caption{The (a) 2b-RQ, (b) 2b-LQU, and (c) 2b-LQ quantizer at $P_\text{e}=0.01$, along with the PDF for the raw data.}
\label{Figure4}
\end{figure}
\begin{figure}[htbp]
\centering
\includegraphics[width=\linewidth]{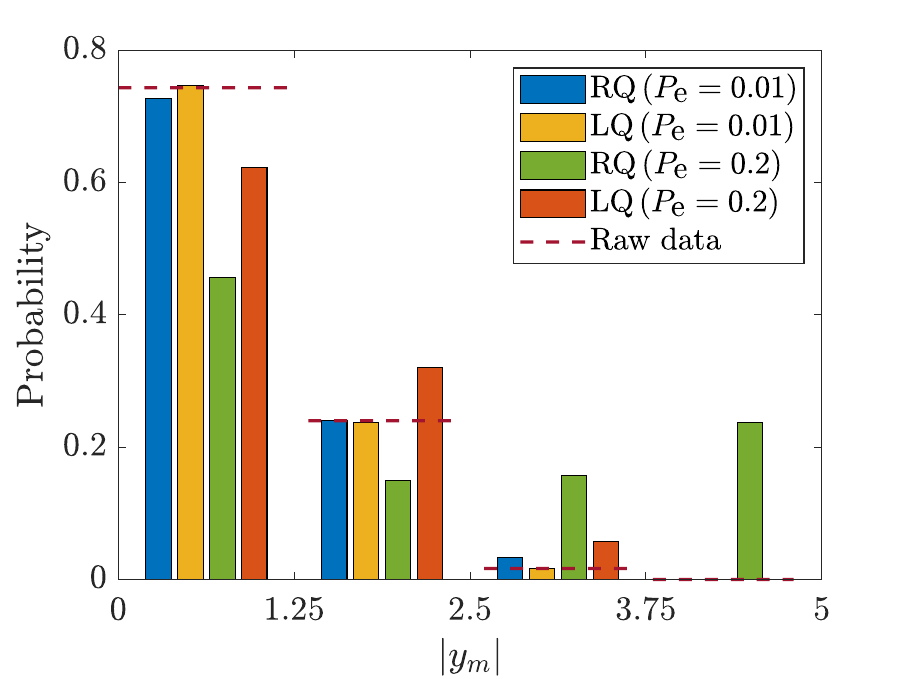}
\caption{Raw data distribution after quantization with 2b-RQ and 2b-LQ respectively and following passing through a BSC with $P_\text{e}\in\left\lbrace 0.01,0.2\right\rbrace$.}
\label{Figure5}
\end{figure}

Figure\,\ref{Figure4} depicts the 2b-RQ, 2b-LQU and 2b-LQ quantizer at $ P_\text{e}=0.01$.  The blue curve in the figure represents the PDF of the raw data $y_m$, while the black dashed line indicates the position of the quantization threshold. We can observe from Fig.\,\ref{Figure4} that RQ is more probable with a probability of $2P_\text{e}(1-P_\text{e})$ to transmit the data with the highest probability of event encoded as the binary code ``10" to the data with the lowest likelihood of occurrence encoded as ``00" or ``11". For LQ, the high-probability  region is encoded as ``00", while the low-probability region is encoded as ``11", resulting in a significantly greater Hamming distance between them. As a consequence, the probability of a communication error occurring is $P_\text{e}^2$ which is notably smaller than that of RQ for any $P_\text{e}<2/3$. 

Figure\,\ref{Figure5} illustrates the distribution of the raw data following quantization with 2b-RQ and 2b-LQ, respectively, and subsequent transmission through a BSC with $P_\text{e}\in\left\lbrace 0.01,0.2\right\rbrace$. To enable comparison, the data is restructured into four uniform intervals (defined by 2b-LQU) after passing through the BSC, based on their binary codes. From Fig.\,\ref{Figure5}, it is evident that  low $P_\text{e}$ values can strike low erroneous data transfer probability for both LQ and RQ. However, when $P_\text{e}=0.2$, a greater amount of data is transferred from the high-probability region (encoded as ``10") to the low-probability region (encoded as ``00" or ``11") of RQ.
\section{Simulation Results}\label{Section5}
The length of sparse signals is fixed at $N = 1000$ in all runs. Following the procedure described in \cite{wang2019detection}, the elements in the  linear compression operators $\pmb{h}_m$ for $m = 1, 2,..., M$ are sampled from an i.i.d. standard normal distribution and then normalized to 1. The SNR at each local sensor is given by 
\begin{align}\label{Eq48}
	\text{SNR}\triangleq\frac{\frac{1}{N}\mathbb{E}\left[|\!|\pmb{s}_l|\!|_2^2\right]}{\mathbb{E}\left[w_l^2\right]}
	=\frac{p\sigma_0^2}{\sigma_w^2}.
\end{align}

\begin{figure*}[htbp]
\centering
\includegraphics[width=\linewidth]{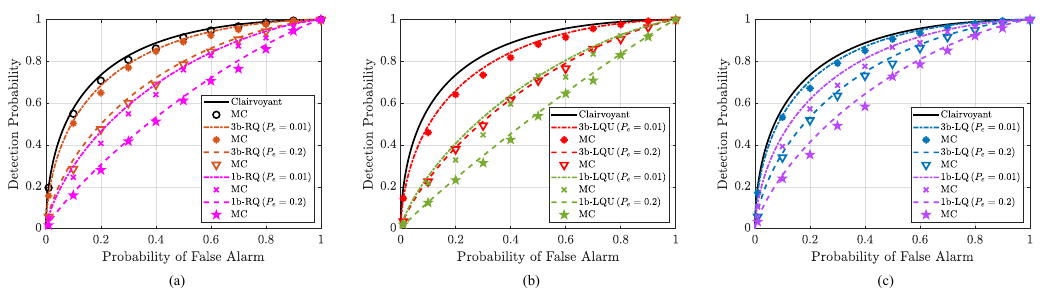}
\caption{ROC curves of the proposed $q$-bit detectors and the Clairvoyant detector, where $q\in\left\lbrace 1, 3\right\rbrace$, $M=300$, $\sigma_s^2=4$, $\sigma_w^2=1$, $p=0.03$, and $P_\text{e}\in\left\lbrace 0.01,0.2\right\rbrace$. (a) RQ, (b) LQU, (c) LQ.}
\label{Figure6}
\end{figure*}

The receiver operating characteristic (ROC) curves of the proposed detectors with varied $q$ and $P_\text{e}$ values are shown in Fig.\,\ref{Figure6}, where $M=300$, $\sigma_0^2=4$, $\sigma_w^2=1$, $p=0.03$, and $P_\text{e}\in\left\lbrace 0.01,0.2\right\rbrace$. As a result, the corresponding $\text{SNR}$ equals to  $-9.2\,\text{dB}$. The lines in Fig.\,\ref{Figure6} reflect theoretical performance, whereas the markers represent the performance through $5000$ Monte Carlo (MC) trials. Figure\,\ref{Figure6} also demonstrates the performance of the Clairvoyant detector as a benchmark. As seen in Fig.\,\ref{Figure6}, the MC results are in accordance with the theoretical performance, and the detection performance improves as the bit depth increases. When the crossover probability is relatively small, the detection performance of a $3$-bit quantizer is similar to that of the Clairvoyant detector. However, as $P_\text{e}$ increases, the performance of all quantized detectors degrades. It can also be observed that LQ  consistently performs better than RQ and LQU, which is in accordance with the FI results in Fig.\,\ref{Figure3}. This is because the LQ approach (illustrated in Figs.\,\ref{Figure4} and \ref{Figure5}) is less likely to incorrectly transfer data from high occurrence probability regions to low occurrence probability regions, resulting in less information loss, which is  particularly evident when $q$ is small.
\begin{figure}[!htbp]
\centering
\includegraphics[width=\linewidth]{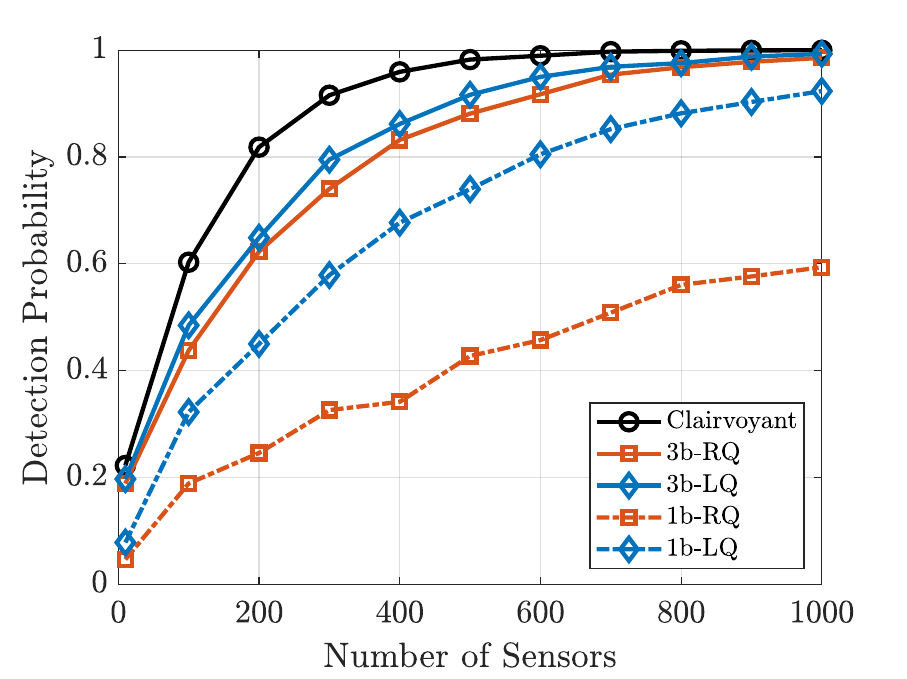}
\caption{Detection probability versus the  number of sensors for the proposed $q$-bit detectors and the Clairvoyant detector, where $q\in\left\lbrace 1, 3\right\rbrace$, $\sigma_0^2=8$, $\sigma_w^2=1$, $p=0.03$, $P_\text{FA}=0.1$ and $P_\text{e}=0.1$.}
\label{Figure7}
\end{figure}
\begin{figure}[!htbp]
\centering
\setlength{\abovecaptionskip}{0.cm}
\setlength{\abovecaptionskip}{0.cm}
\setlength{\belowdisplayskip}{3pt}
\includegraphics[width=\linewidth]{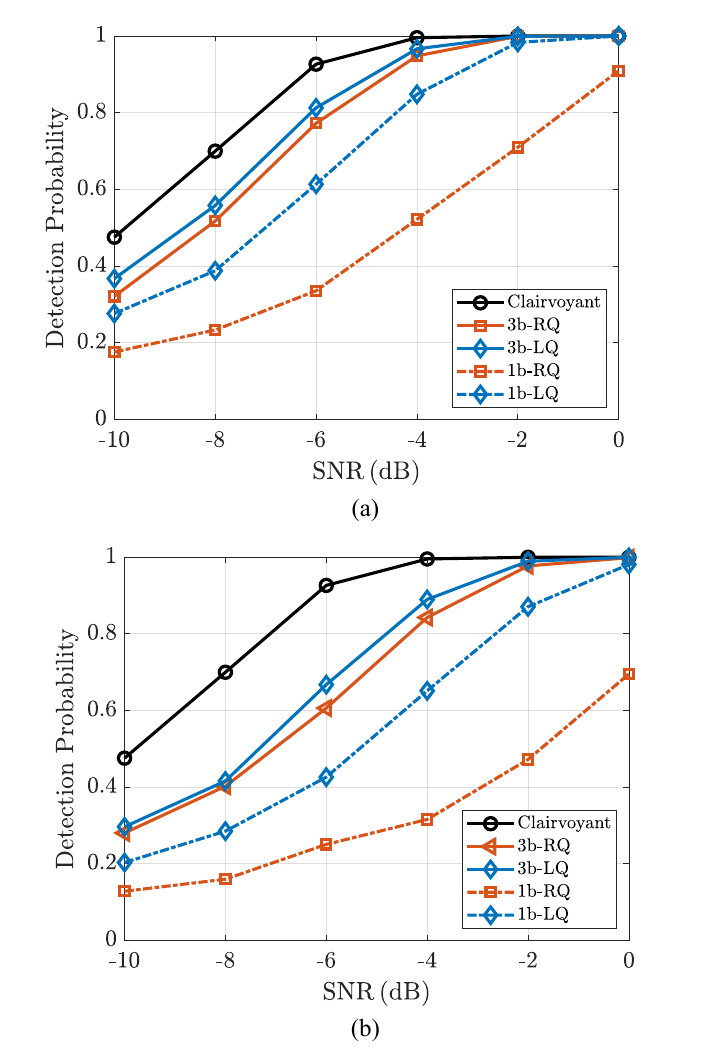}
\caption{Detection probability versus the  SNR for the proposed $q$-bit detectors and the Clairvoyant detector, where $q\in\left\lbrace 1, 3\right\rbrace$, $M=300$, $\sigma_w^2=1$, $p=0.03$, and $P_\text{FA}=0.1$.(a) $P_\text{e}=0.1$, (b) $P_\text{e}=0.2$.}
\label{Figure8}
\end{figure}

In Fig.\,\ref{Figure7}, we plot the detection probabilities versus the number of sensors  for the proposed $q$-bit detectors and the Clairvoyant detector, where $q\in\left\lbrace 1, 3\right\rbrace$, $\sigma_s^2=8$, $\sigma_w^2=1$, $p=0.03$, $P_\text{FA}=0.1$ and $P_\text{e}=0.1$, while the $\text{SNR}$  is $-6.2\,\text{dB}$. The performance of each detector increases by invoking more sensors, as shown in Fig.\,\ref{Figure7}. When $M = 300$, the detection probabilities of 3b-RQ and 3b-LQ are 0.7404 and 0.795, respectively, while the detection probabilities of 1b-RQ and 1b-LQ are 0.3262 and 0.5788, respectively. This further supports the superiority of LQ. On the other hand, when the quantization depth is small, particularly for RQ, the detection probability grows slowly with increasing $M$, even though adding more sensors can enhance the detection performance. This demonstrates the importance of raising the quantization depth in order to improve the detection performance of WSN systems.

The detection performance of the proposed $q$-bit detectors and the Clairvoyant detector with different values of $\text{SNR}$ are investigated in Fig.\,\ref{Figure8}, where $q\in\left\lbrace 1, 3\right\rbrace$, $M=300$, $\sigma_w^2=1$, $p=0.03$, $P_\text{FA}=0.1$, and $P_\text{e}\in\left\lbrace 0.01,0.2\right\rbrace$.  Bearing in mind that the $\text{SNR}$ described in \eqref{Eq48} takes both the sparsity degree and the $\text{SNR}$ of each non-zero component of the sparse signal into consideration. As shown in Fig.\,\ref{Figure8}, all detectors perform better as $\text{SNR}$ rises. When $P_\text{e}=0.1$, 3b-RQ and 3b-LQ may obtain a near-Clairvoyant detection probability at $\text{SNR}=-4\,\text{dB}$. The performance of each quantized detector diminishes as $P_\text{e}$ grows, as shown by Fig.\,\ref{Figure8}. Nevertheless, under the scenario of $P_\text{e}=0.2$, both 3b-RQ and 3b-LQ can attain the same detection probability ($P_\text{D}=1$) as the clairvoyant  detector at $\text{SNR}=0\,\text{dB}$. Although the detection probability of 1b-LQ is $0.9812$ at $0\,\text{dB}$, in the SNR interval of $[-10,-2]\,\text{dB}$, its detection probability  is lower by $0.1\sim0.2$ than that of 3b-LQ, while the detection probability of 1b-RQ is only $0.6952$. Based on the aforementioned findings, it is clear that multi-bit detectors are required in non-ideal channels.

\begin{figure}[htbp]
\centering
\includegraphics[width=\linewidth]{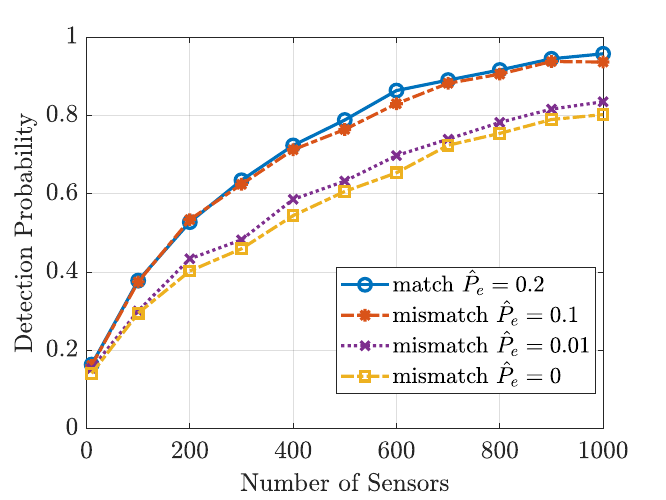}
\caption{Detection probability versus the  number of sensors for the proposed $3$-bit LQ detector, with an actual crossover probability $P_\text{e}=0.2$, estimated crossover probabilities $\hat{P}_\text{e}\in\left\lbrace 0,0.01,0.1,0.2\right\rbrace$, $\sigma_0^2=8$, $\sigma_w^2=1$, $p=0.03$, and $P_\text{FA}=0.1$.}
\label{Figure9}
\end{figure}

We subsequently evaluate the impact of inaccurate $P_\text{e}$ on the detection performance of the proposed $q$-bit detectors. Figure\,\ref{Figure9}  illustrates the detection probability curves across various sensor counts, using a 3-bit LQ detector as a representative case, with an actual crossover probability $P_\text{e}=0.2$ and estimated crossover probabilities $\hat{P}_\text{e}\in\left\lbrace 0,0.01,0.1,0.2\right\rbrace$. From Fig.\,\ref{Figure9}, it is evident that reduction in the precision of crossover probability estimates can lead to significant degradation in detection performance: the greater the estimation error, the more substantial  the loss in performance.
\section{Summary and Conclusions}\label{Section6}
This paper examined the problem of multi-bit distributed detection over error-prone channels for sparse stochastic signals. We developed two multi-bit LMPT detectors at the FC that fuse the quantized raw observations and local likelihood ratios, respectively. The asymptotic detection performance of the proposed quantized detectors was analyzed. Closed-form expressions for the detection and false alarm probabilities were derived, and then confirmed by Monte Carlo simulations. We further optimized the quantization thresholds for both RQ and LQ by maximizing their associated FI.
A comprehensive performance analysis was conducted to characterize and compensate for the quantization-induced information loss of proposed detectors. A comparison was also provided to explain the performance difference between RQ and LQ. The theoretical analysis was validated by simulation findings, confirming the importance of increasing the quantization depth to improve WSN detection performance in error-prone channels.

\appendices
\section{Proof of \eqref{Eq12}}\label{Appendices1}
\begin{proof}[\unskip\nopunct]
The logarithm of \eqref{Eq10} can be calculated as:
\begin{align}\label{Eq49}
 \ln {P}(\pmb{d}^\text{rq}|\mathcal{H}_1;p)=\sum_{m=1}^{M}\sum_{i=1}^{2^q}I(\pmb{d}_m^\text{rq},\pmb{z}_{m,i}^\text{rq})\ln\xi_{m,i}^\text{rq},
 \end{align}
 where 
 \begin{align}\label{Eq50}
 \xi_{m,i}^\text{rq}\triangleq\sum_{j=1}^{2^q}G(q,P_\text{e},D_{m,i,j}^\text{rq})Q_{m,j}^\text{rq}(p).
 \end{align}
 Taking the derivative of \eqref{Eq49} with respect to $p$ results in
\begin{align}\label{Eq51}
\frac{\partial\ln {P}(\pmb{d}^\text{rq}|\mathcal{H}_1;p)}{\partial p}=\sum_{m=1}^{M}\sum_{i=1}^{2^q}I(\pmb{d}_m^\text{rq},\pmb{z}_{m,i}^\text{rq})\frac{\dot{\xi}_{m,i}^\text{rq}}{\xi_{m,i}^\text{rq}},
\end{align}
where $\dot{\xi}_{m,i}^\text{rq}$ represents the first derivative of $\xi_{m,i}^\text{rq}$, which can be further derived as
\begin{align}\label{Eq52}
\dot{\xi}_{m,i}^\text{rq}&=\frac{\sigma_0^2}{2}\frac{|\!|\pmb{h}_m|\!|_2^2}{\sigma_m^{2}(p,\sigma_0^2)}\big \lbrace 
 -G(q,\!P_\text{e},\!D_{m,i,1}^\text{rq})\bar{\tau}_{m,1}^\text{rq}\Psi\left(\bar{\tau}_{m,1}^\text{rq}\right)\nonumber\\
&+G(q,P_\text{e},D_{m,i,2}^\text{rq})\left[\bar{\tau}_{m,1}^\text{rq}\Psi\left(\bar{\tau}_{m,1}^\text{rq}\right)-\bar{\tau}_{m,2}^\text{rq}\Psi\left(\bar{\tau}_{m,2}^\text{rq}\right)\right]\nonumber\\
&+\cdots+G(q,P_\text{e},D_{m,i,2^q}^\text{rq})\left[\bar{\tau}_{m,2^q-1}^\text{rq}\Psi\left(\bar{\tau}_{m,2^q-1}^\text{rq}\right)\right]\big\rbrace\nonumber\\
&=\frac{\sigma_0^2}{2}\frac{|\!|\pmb{h}_m|\!|_2^2}{\sigma_m^{3 }(p,\sigma_0^2)}\sum_{j=1}^{2^q}G(q,P_\text{e},D_{m,i,j}^\text{rq})F_{m,j}^\text{rq}(p),
\end{align}
where 
\begin{align}\label{Eq53}
\bar{\tau}_{m,2^q-1}^\text{rq}\!\triangleq{\tau_{m,2^q-1}^\text{rq}}/{\sigma_m(p,\sigma_0^2)},
\end{align}
and
\begin{align}\label{Eq54}
\Psi(x)=1/\sqrt{2\pi}\exp(-x^2/2).
\end{align}
Substituting \eqref{Eq50} and \eqref{Eq52} into \eqref{Eq51} yields \eqref{Eq12}. This completes the proof of  \eqref{Eq12}.
\end{proof}
\section{Proof of \eqref{Eq14}}\label{Appendices2}
\begin{proof}[\unskip\nopunct]
Taking the second derivative of the logarithm in \eqref{Eq10} with respect to $p$ leads to
\begin{multline}\label{Eq55}
\frac{\partial^2\ln {P}(\pmb{d}^\text{rq}|\mathcal{H}_1;p)}{\partial p^2}\\
=\sum_{m=1}^{M}\sum_{i=1}^{2^q}I(\pmb{d}_m^\text{rq},\pmb{z}_{m,i}^\text{rq})\left[\frac{\ddot{\xi}_{m,i}^\text{rq}}{\xi_{m,i}^\text{rq}}-\left(\frac{\dot{\xi}_{m,i}^\text{rq}}{\xi_{m,i}^\text{rq}}\right)^{2}\right],
\end{multline}
with $\ddot{\xi}_{m,i}^\text{rq}$ being the second derivative of ${\xi}_{m,i}^\text{rq}$, which we divide into two terms, i.e., 
\begin{align}\label{Eq56}
\nu_1=-\frac{3\sigma_0^4}{4}\frac{|\!|\pmb{h}_m|\!|_2^4}{\sigma_m^{5}(p,\sigma_0^2)}\sum_{j=1}^{2^q}G(q,P_\text{e},D_{m,i,j}^\text{rq})F_{m,j}^\text{rq}(p),
\end{align}
and
\begin{align}\label{Eq57}
\nu_2&=\frac{\sigma_0^4}{4}\frac{|\!|\pmb{h}_m|\!|_2^4}{\sigma_m^{4}(p,\sigma_0^2)}\big \lbrace 
 -G(q,\!P_\text{e},\!D_{m,i,1}^\text{rq})(\bar{\tau}_{m,1}^\text{rq})^2\Omega\left(\bar{\tau}_{m,1}^\text{rq}\right)\nonumber\\
&+G(q,\!P_\text{e},\!D_{m,i,2}^\text{rq})\left[(\bar{\tau}_{m,1}^\text{rq})^2\Omega\left(\bar{\tau}_{m,1}^\text{rq}\right)-(\bar{\tau}_{m,2}^\text{rq})^2\Omega\left(\bar{\tau}_{m,2}^\text{rq}\right)\right]\nonumber\\
&\!+\!\cdots\!+\!G(q,\!P_\text{e},\!D_{m,i,2^q}^\text{rq})\left[(\bar{\tau}_{m,2^q-1}^\text{rq})^2\Omega\left(\bar{\tau}_{m,2^q-1}^\text{rq}\right)\right]\big\rbrace,
\end{align}
where
\begin{align}\label{Eq58}
\Omega(x)=-x/\sqrt{2\pi}\exp(-x^2/2).
\end{align}
Define
\begin{align}\label{Eq59}
J(\pmb{d}_m^\text{rq};p)\triangleq\!\sum_{i=1}^{2^q}I(\pmb{d}_m^\text{rq},\pmb{z}_{m,i}^\text{rq})\!\left[\frac{\ddot{\xi}_{m,i}^\text{rq}}{\xi_{m,i}^\text{rq}}-\left(\frac{\dot{\xi}_{m,i}^\text{rq}}{\xi_{m,i}^\text{rq}}\right)^{2}\right].
\end{align}
The Fisher information in \eqref{Eq11} can be calculated as
\begin{align}\label{Eq60}
\text{FI}_q^\text{rq}(p)
&\triangleq\!-E\left[\!\frac{\partial^2\ln P(\pmb{d}^\text{rq}|\mathcal{H}_1;p)}{\partial p^2}\right]\nonumber\\
&=-\sum_{m=1}^{M}\sum_{i=1}^{2^q}J(\pmb{d}_m^\text{rq}=\pmb{z}_{m,i}^\text{rq};p){\xi}_{m,i}^\text{rq}\nonumber\\
&=-\sum_{m=1}^{M}\sum_{i=1}^{2^q}\bigg[\ddot{\xi}_{m,i}^\text{rq}-\frac{\big(\dot{\xi}_{m,i}^\text{rq}\big)^{2}}{\xi_{m,i}^\text{rq}}\bigg].
\end{align}
Given $\sum_{i=1}^{2^q}G(q,\!P_\text{e},\!D_{m,i,j}^\text{rq})=1$, we can deduce that
\begin{align}\label{Eq61}
\sum_{i=1}^{2^q}\nu_1
&=-\frac{3\sigma_0^4}{4}\frac{|\!|\pmb{h}_m|\!|_2^4}{\sigma_m^{4}(p,\sigma_0^2)}\big[
 -\bar{\tau}_{m,1}^\text{rq}\Psi\left(\bar{\tau}_{m,1}^\text{rq}\right)\nonumber\\
&+\bar{\tau}_{m,1}^\text{rq}\Psi\left(\bar{\tau}_{m,1}^\text{rq}\right)-\bar{\tau}_{m,2}^\text{rq}\Psi\left(\bar{\tau}_{m,2}^\text{rq}\right)\nonumber\\
&+\cdots+\bar{\tau}_{m,2^q-1}^\text{rq}\Psi\left(\bar{\tau}_{m,2^q-1}^\text{rq}\right)\big]
=0,
\end{align}
and 
\begin{align}\label{Eq62}
\sum_{i=1}^{2^q}\nu_2
&=\frac{\sigma_0^4}{4}\frac{|\!|\pmb{h}_m|\!|_2^4}{\sigma_m^{4}(p,\sigma_0^2)}\big[
 -(\bar{\tau}_{m,1}^\text{rq})^2\Omega\left(\bar{\tau}_{m,1}^\text{rq}\right)\nonumber\\
&+(\bar{\tau}_{m,1}^\text{rq})^2\Omega\left(\bar{\tau}_{m,1}^\text{rq}\right)-(\bar{\tau}_{m,2}^\text{rq})^2\Omega\left(\bar{\tau}_{m,2}^\text{rq}\right)\nonumber\\
&+\!\cdots+\!(\bar{\tau}_{m,2^q-1}^\text{rq})^2\Omega\left(\bar{\tau}_{m,2^q-1}^\text{rq}\right)\big]=0.
\end{align}
Combining \eqref{Eq61} and \eqref{Eq62}, we can conclude that 
\begin{align}\label{Eq63}
\sum_{i=1}^{2^q}\ddot{\xi}_{m,i}^\text{rq}=0.
\end{align}
Substituting \eqref{Eq50} \eqref{Eq52} and \eqref{Eq63} into \eqref{Eq60} yields \eqref{Eq14}. This completes the proof of \eqref{Eq14}.
\end{proof}
\section{Proof of \eqref{Eq24}}\label{Appendices3}
\begin{proof}[\unskip\nopunct]
The logarithm of \eqref{Eq24} can be calculated as:
\begin{align}\label{Eq64}
 \ln P(\pmb{d}^\text{lq}|\mathcal{H}_1;p)=\sum_{m=1}^{M}\sum_{i=1}^{2^q}I(\pmb{d}_m^\text{lq},\pmb{z}_{m,i}^\text{rq})\ln\xi_{m,i}^\text{lq},
 \end{align}
 where 
 \begin{align}\label{Eq65}
 \xi_{m,i}^\text{lq}\triangleq\sum_{j=1}^{2^q}2G(q,P_\text{e},D_{m,i,j}^\text{lq})Q_{m,j}^\text{lq}(p).
 \end{align}
 Taking the derivative of \eqref{Eq64} with respect to $p$ results in
\begin{align}\label{Eq66}
\frac{\partial\ln P(\pmb{d}^\text{lq}|\mathcal{H}_1;p)}{\partial p}=\sum_{m=1}^{M}\sum_{i=1}^{2^q}I(\pmb{d}_m^\text{lq},\pmb{z}_{m,i}^\text{lq})\frac{\dot{\xi}_{m,i}^\text{lq}}{\xi_{m,i}^\text{lq}},
\end{align}
where $\dot{\xi}_{m,i}^\text{lq}$ represents the first derivative of $\xi_{m,i}^\text{lq}$, which can be further derived as
\begin{align}\label{Eq67}
\dot{\xi}_{m,i}^\text{lq}&=\frac{\sigma_0^2|\!|\pmb{h}_m|\!|_2^2}{\sigma_m^{2}(p,\sigma_0^2)}\big \lbrace 
 -G(q,\!P_\text{e},\!D_{m,i,1}^\text{lq})\bar{\tau}_{m,1}^\text{lq}\Psi\left(\bar{\tau}_{m,1}^\text{lq}\right)\nonumber\\
&+G(q,P_\text{e},D_{m,i,2}^\text{lq})\left[\bar{\tau}_{m,1}^\text{lq}\Psi\left(\bar{\tau}_{m,1}^\text{lq}\right)-\bar{\tau}_{m,2}^\text{lq}\Psi\left(\bar{\tau}_{m,2}^\text{lq}\right)\right]\nonumber\\
&+\cdots+G(q,P_\text{e},D_{m,i,2^q}^\text{lq})\left[\bar{\tau}_{m,2^q-1}^\text{lq}\Psi\left(\bar{\tau}_{m,2^q-1}^\text{lq}\right)\right]\big\rbrace\nonumber\\
&=\frac{\sigma_0^2|\!|\pmb{h}_m|\!|_2^2}{\sigma_m^{3 }(p,\sigma_0^2)}\sum_{j=1}^{2^q}G(q,P_\text{e},D_{m,i,j}^\text{lq})F_{m,j}^\text{lq}(p).
\end{align}
Substituting \eqref{Eq65} and \eqref{Eq67} into \eqref{Eq66} yields \eqref{Eq24}. This completes the proof of  \eqref{Eq24}.
\end{proof}
\section{Proof of \eqref{Eq26}}\label{Appendices4}
\begin{proof}[\unskip\nopunct]
Taking the second derivative of the logarithm in \eqref{Eq21} with respect to $p$ leads to
\begin{multline}\label{Eq68}
\frac{\partial^2\ln P(\pmb{d}^\text{lq}|\mathcal{H}_1;p)}{\partial p^2}\\
=\sum_{m=1}^{M}\sum_{i=1}^{2^q}I(\pmb{d}_m^\text{lq},\pmb{z}_{m,i}^\text{lq})\left[\frac{\ddot{\xi}_{m,i}^\text{lq}}{\xi_{m,i}^\text{lq}}-\left(\frac{\dot{\xi}_{m,i}^\text{lq}}{\xi_{m,i}^\text{lq}}\right)^{2}\right],
\end{multline}
where $\ddot{\xi}_{m,i}^\text{lq}$ represents the second derivative of ${\xi}_{m,i}^\text{lq}$. 
The Fisher information for LQ fusion can be calculated as
\begin{align}\label{Eq69}
\text{FI}_q^\text{lq}(p)
&\triangleq\!-E\left[\!\frac{\partial^2\ln P(\pmb{d}^\text{lq}|\mathcal{H}_1;p)}{\partial p^2}\right]\nonumber\\
&=-\sum_{m=1}^{M}\sum_{i=1}^{2^q}J(\pmb{d}_m^\text{lq}=\pmb{z}_{m,i}^\text{lq};p){\xi}_{m,i}^\text{lq}\nonumber\\
&=-\sum_{m=1}^{M}\sum_{i=1}^{2^q}\bigg[\ddot{\xi}_{m,i}^\text{lq}-\frac{\big(\dot{\xi}_{m,i}^\text{lq}\big)^{2}}{\xi_{m,i}^\text{lq}}\bigg].
\end{align}
By employing a similar derivation approach to \eqref{Eq56}$\sim$\eqref{Eq62}, we can conclude that
\begin{align}\label{Eq70}
\sum_{i=1}^{2^q}\ddot{\xi}_{m,i}^\text{lq}=0.
\end{align}
Substituting \eqref{Eq65} \eqref{Eq67} and \eqref{Eq70} into \eqref{Eq69} yields \eqref{Eq26}. This completes the proof of \eqref{Eq26}.
\end{proof}
\renewcommand{\refname}{References}
\mbox{} 
\nocite{*}
\bibliographystyle{IEEEtran}
\bibliography{reference.bib}

\begin{thebibliography}{10}
\providecommand{\url}[1]{#1}
\csname url@samestyle\endcsname
\providecommand{\newblock}{\relax}
\providecommand{\bibinfo}[2]{#2}
\providecommand{\BIBentrySTDinterwordspacing}{\spaceskip=0pt\relax}
\providecommand{\BIBentryALTinterwordstretchfactor}{4}
\providecommand{\BIBentryALTinterwordspacing}{\spaceskip=\fontdimen2\font plus
\BIBentryALTinterwordstretchfactor\fontdimen3\font minus \fontdimen4\font\relax}
\providecommand{\BIBforeignlanguage}[2]{{%
\expandafter\ifx\csname l@#1\endcsname\relax
\typeout{** WARNING: IEEEtran.bst: No hyphenation pattern has been}%
\typeout{** loaded for the language `#1'. Using the pattern for}%
\typeout{** the default language instead.}%
\else
\language=\csname l@#1\endcsname
\fi
#2}}
\providecommand{\BIBdecl}{\relax}
\BIBdecl

\bibitem{kay1998fundamentals}
S.~M. Kay, \emph{Fundamentals of statistical signal processing: Detection Theory}.\hskip 1em plus 0.5em minus 0.4em\relax Upper Saddle River, NJ: Prentice-Hall, 1998.

\bibitem{wang2018particle}
D.~Wang, D.~Tan, and L.~Liu, ``Particle swarm optimization algorithm: an overview,'' \emph{Soft computing}, vol.~22, no.~2, pp. 387--408, 2018.

\bibitem{djordjevic2019physical}
I.~B. Djordjevic, \emph{Physical-layer security and quantum key distribution}.\hskip 1em plus 0.5em minus 0.4em\relax Tucson, AZ: Springer, 2019.

\bibitem{daneshgaran2014ldpc}
F.~Daneshgaran, M.~Mondin, and I.~Bari, ``{LDPC} coding for {QKD} at higher photon flux levels based on spatial entanglement of twin beams in {PDC},'' in \emph{Journal of Physics: Conference Series}, vol. 497, no.~1.\hskip 1em plus 0.5em minus 0.4em\relax IOP Publishing, 2014, p. 012037.

\bibitem{wang2022proposal}
X.~Wang, D.~Zhu, G.~Li, X.-P. Zhang, and Y.~He, ``Proposal-copula-based fusion of spaceborne and airborne {SAR} images for ship target detection,'' \emph{Information Fusion}, vol.~77, pp. 247--260, 2022.

\bibitem{li2024comic}
C.~Li, G.~Li, Z.~Wang, X.~Wang, and P.~K. Varshney, ``{COMIC}: An unsupervised change detection method for heterogeneous remote sensing images based on copula mixtures and cycle-consistent adversarial networks,'' \emph{Information Fusion}, vol. 106, p. 102240, 2024.

\bibitem{Zayyani2016iterative}
M.~Korki, J.~Zhang, C.~Zhang, and H.~Zayyani, ``Iterative {Bayesian} reconstruction of non-{IID} block-sparse signals,'' \emph{IEEE Transactions on Signal Processing}, vol.~64, no.~13, pp. 3297--3307, Jul. 2016.

\bibitem{donoho2006compressed}
D.~L. Donoho, ``Compressed sensing,'' \emph{IEEE Transactions on information theory}, vol.~52, no.~4, pp. 1289--1306, Apr. 2006.

\bibitem{sui2023space}
Z.~Sui, H.~Zhang, S.~Sun, L.-L. Yang, and L.~Hanzo, ``Space-time shift keying aided {OTFS} modulation for orthogonal multiple access,'' \emph{IEEE Transactions on Communications}, vol.~71, no.~12, pp. 7393--7408, Sep. 2023.

\bibitem{yang2023hybrid}
S.~Yang, Y.~Lai, A.~Jakobsson, and W.~Yi, ``Hybrid quantized signal detection with a bandwidth-constrained distributed radar system,'' \emph{IEEE Transactions on Aerospace and Electronic Systems}, vol.~59, no.~6, pp. 7835--7850, Dec. 2023.

\bibitem{sui2023performance}
Z.~Sui, S.~Yan, H.~Zhang, S.~Sun, Y.~Zeng, L.-L. Yang, and L.~Hanzo, ``Performance analysis and approximate message passing detection of orthogonal time sequency multiplexing modulation,'' \emph{IEEE Transactions on Wireless Communications}, vol.~23, no.~3, pp. 1913--1928, Mar. 2024.

\bibitem{zhang2023direct}
G.~Zhang, W.~Yi, P.~K. Varshney, and L.~Kong, ``Direct target localization with quantized measurements in non-coherent distributed {MIMO} radar systems,'' \emph{IEEE Transactions on Geoscience and Remote Sensing}, early access, 2023, doi:10.1109/TGRS.2023.3267499.

\bibitem{feng2023compressive}
Y.~Feng, A.~Taya, Y.~Nishiyama, K.~Sezaki, and J.~Liu, ``Compressive detection of stochastic sparse signals with unknown sparsity degree,'' \emph{IEEE Signal Processing Letters}, vol.~30, pp. 1482--1486, 2023.

\bibitem{tabella2023time}
G.~Tabella, D.~Ciuonzo, Y.~Yilmaz, X.~Wang, and P.~S. Rossi, ``Time-aware distributed sequential detection of gas dispersion via wireless sensor networks,'' \emph{IEEE Transactions on Signal and Information Processing over Networks}, vol.~9, pp. 721--735, 2023.

\bibitem{cheng2022dynamic}
M.~Cheng, Q.~Guan, F.~Ji, J.~Cheng, and Y.~Chen, ``Dynamic-detection-based trajectory planning for autonomous underwater vehicle to collect data from underwater sensors,'' \emph{IEEE Internet of Things Journal}, vol.~9, no.~15, pp. 13\,168--13\,178, Aug. 2022.

\bibitem{wei2021reliable}
X.~Wei, H.~Guo, X.~Wang, X.~Wang, and M.~Qiu, ``Reliable data collection techniques in underwater wireless sensor networks: A survey,'' \emph{IEEE Communications Surveys \& Tutorials}, vol.~24, no.~1, pp. 404--431, 2021.

\bibitem{zhu2022adaptive}
F.~Zhu, J.~H. Park, and L.~Peng, ``Adaptive event-triggered quantized communication-based distributed estimation over sensor networks with {Semi-Markovian} switching topologies,'' \emph{IEEE Transactions on Signal and Information Processing over Networks}, vol.~8, pp. 258--272, 2022.

\bibitem{wang2021target}
Z.~Wang, Q.~He, and R.~S. Blum, ``Target detection using quantized cloud {MIMO} radar measurements,'' \emph{IEEE Transactions on Signal Processing}, vol.~70, pp. 1--16, 2021.

\bibitem{zang2022fast}
Y.~Zang and H.~Zhu, ``Fast and optimal joint decision and estimation by quantized data via noisy channels of sensor networks,'' \emph{Signal Processing}, vol. 195, p. 108481, 2022.

\bibitem{Mohammadi2022generalized}
A.~Mohammadi, D.~Ciuonzo, A.~Khazaee, and P.~S. Rossi, ``Generalized locally most powerful tests for distributed sparse signal detection,'' \emph{IEEE Transactions on Signal and Information Processing over Networks}, vol.~8, pp. 528--542, 2022.

\bibitem{maya2021fully}
J.~A. Maya and L.~R. Vega, ``On fully-distributed composite tests with general parametric data distributions in sensor networks,'' \emph{IEEE Transactions on Signal and Information Processing over Networks}, vol.~7, pp. 509--521, 2021.

\bibitem{hu2020decentralized}
L.~Hu, X.~Wang, and S.~Wang, ``Decentralized underwater target detection and localization,'' \emph{IEEE Sensors Journal}, vol.~21, no.~2, pp. 2385--2399, 2020.

\bibitem{van2021distributed}
R.~Van~Rompaey and M.~Moonen, ``Distributed adaptive signal estimation in wireless sensor networks with partial prior knowledge of the desired sources steering matrix,'' \emph{IEEE Transactions on Signal and Information Processing over Networks}, vol.~7, pp. 478--492, 2021.

\bibitem{ciuonzo2021distributed}
D.~Ciuonzo, P.~S. Rossi, and P.~K. Varshney, ``Distributed detection in wireless sensor networks under multiplicative fading via generalized score tests,'' \emph{IEEE Internet of Things Journal}, vol.~8, no.~11, pp. 9059--9071, Jun 2021.

\bibitem{cheng2021multi}
X.~Cheng, D.~Ciuonzo, P.~S. Rossi, X.~Wang, and W.~Wang, ``Multi-bit {\&} sequential decentralized detection of a noncooperative moving target through a generalized {Rao} test,'' \emph{IEEE Transactions on Signal and Information Processing over Networks}, vol.~7, pp. 740--753, 2021.

\bibitem{liang2020distributed}
T.~Liang, Y.~Lin, L.~Shi, J.~Li, Y.~Zhang, and Y.~Qian, ``Distributed vehicle tracking in wireless sensor network: A fully decentralized multiagent reinforcement learning approach,'' \emph{IEEE Sensors Letters}, vol.~5, no.~1, pp. 1--4, Jan. 2020.

\bibitem{zhang2020Distributed}
S.~Zhang, P.~Khanduri, and P.~K. Varshney, ``Distributed sequential detection: Dependent observations and imperfect communication,'' \emph{IEEE Transactions on Signal Processing}, vol.~68, pp. 830--842, 2020.

\bibitem{li2020falsified}
C.~Li, G.~Li, and P.~K. Varshney, ``Distributed detection of sparse signals with physical layer secrecy constraints: A falsified censoring strategy,'' \emph{IEEE Transactions on Signal Processing}, vol.~68, pp. 6040--6054, 2020.

\bibitem{li2020tree}
------, ``Distributed detection of sparse stochastic signals with 1-bit data in tree-structured sensor networks,'' \emph{IEEE Transactions on Signal Processing}, vol.~68, pp. 2963--2976, 2020.

\bibitem{li2020censoring}
------, ``Distributed detection of sparse signals with censoring sensors via locally most powerful test,'' \emph{IEEE Signal Processing Letters}, vol.~27, pp. 346--350, 2020.

\bibitem{ciuonzo2020bandwidth}
D.~Ciuonzo, S.~H. Javadi, A.~Mohammadi, and P.~S. Rossi, ``Bandwidth-constrained decentralized detection of an unknown vector signal via multisensor fusion,'' \emph{IEEE Transactions on Signal and Information Processing over Networks}, vol.~6, pp. 744--758, 2020.

\bibitem{cheng2019multibit}
X.~Cheng, D.~Ciuonzo, and P.~S. Rossi, ``Multibit decentralized detection through fusing smart and dumb sensors based on {Rao} test,'' \emph{IEEE Transactions on Aerospace and Electronic Systems}, vol.~56, no.~2, pp. 1391--1405, Apr. 2020.

\bibitem{luo2019distributed}
J.~Luo, J.~Ni, and Z.~Liu, ``Distributed decision fusion under nonideal communication channels with adaptive topology,'' \emph{Information Fusion}, vol.~45, pp. 190--201, 2019.

\bibitem{wang2019detection}
X.~Wang, G.~Li, and P.~K. Varshney, ``Detection of sparse stochastic signals with quantized measurements in sensor networks,'' \emph{IEEE Transactions on Signal Processing}, vol.~67, no.~8, pp. 2210--2220, Apr. 2019.

\bibitem{wang2019spl}
------, ``Distributed detection of weak signals from one-bit measurements under observation model uncertainties,'' \emph{IEEE Signal Processing Letters}, vol.~26, no.~3, pp. 415--419, Mar. 2019.

\bibitem{wang2019gg}
X.~Wang, G.~Li, C.~Quan, and P.~K. Varshney, ``Distributed detection of sparse stochastic signals with quantized measurements: The generalized {Gaussian} case,'' \emph{IEEE Transactions on Signal Processing}, vol.~67, no.~18, pp. 4886--4898, Sep. 2019.

\bibitem{li2019lr}
C.~Li, Y.~He, X.~Wang, G.~Li, and P.~K. Varshney, ``Distributed detection of sparse stochastic signals via fusion of 1-bit local likelihood ratios,'' \emph{IEEE Signal Processing Letters}, vol.~26, no.~12, pp. 1738--1742, Dec. 2019.

\bibitem{li2019secure}
C.~Li, G.~Li, B.~Kailkhura, and P.~K. Varshney, ``Secure distributed detection of sparse signals via falsification of local compressive measurements,'' \emph{IEEE Transactions on Signal Processing}, vol.~67, no.~18, pp. 4696--4706, Sep. 2019.

\bibitem{aldalahmeh2019fusion}
S.~A. Aldalahmeh, S.~O. Al-Jazzar, D.~McLernon, S.~A.~R. Zaidi, and M.~Ghogho, ``Fusion rules for distributed detection in clustered wireless sensor networks with imperfect channels,'' \emph{IEEE Transactions on Signal and Information Processing over Networks}, vol.~5, no.~3, pp. 585--597, Sep. 2019.

\bibitem{yan2018feedback}
J.~Yan, Z.~Xu, X.~Luo, C.~Chen, and X.~Guan, ``Feedback-based target localization in underwater sensor networks: A multisensor fusion approach,'' \emph{IEEE Transactions on Signal and Information Processing over Networks}, vol.~5, no.~1, pp. 168--180, Mar. 2018.

\bibitem{wang2018detection}
X.~Wang, G.~Li, and P.~K. Varshney, ``Detection of sparse signals in sensor networks via locally most powerful tests,'' \emph{IEEE Signal Processing Letters}, vol.~25, no.~9, pp. 1418--1422, Sep. 2018.

\bibitem{ciuonzo2017quantizer}
D.~Ciuonzo and P.~S. Rossi, ``Quantizer design for generalized locally optimum detectors in wireless sensor networks,'' \emph{IEEE Wireless Communications Letters}, vol.~7, no.~2, pp. 162--165, Apr. 2017.

\bibitem{hu2018decentralized}
L.~Hu, J.~Zhang, X.~Wang, S.~Wang, and E.~Zhang, ``Decentralized truncated one-sided sequential detection of a noncooperative moving target,'' \emph{IEEE Signal Processing Letters}, vol.~25, no.~10, pp. 1490--1494, Oct. 2018.

\bibitem{ciuonzo2017distributed}
D.~Ciuonzo and P.~S. Rossi, ``Distributed detection of a non-cooperative target via generalized locally-optimum approaches,'' \emph{Information Fusion}, vol.~36, pp. 261--274, 2017.

\bibitem{ciuonzo2017generalized}
D.~Ciuonzo, P.~S. Rossi, and P.~Willett, ``Generalized {Rao} test for decentralized detection of an uncooperative target,'' \emph{IEEE Signal Processing Letters}, vol.~24, no.~5, pp. 678--682, May 2017.

\bibitem{zayyani2016double}
H.~Zayyani, F.~Haddadi, and M.~Korki, ``Double detector for sparse signal detection from one-bit compressed sensing measurements,'' \emph{IEEE Signal Processing Letters}, vol.~23, no.~11, pp. 1637--1641, Nov. 2016.

\bibitem{javadi2016detection}
S.~H. Javadi, ``Detection over sensor networks: a tutorial,'' \emph{IEEE Aerospace and Electronic Systems Magazine}, vol.~31, no.~3, pp. 2--18, Mar. 2016.

\bibitem{braca2015distributed}
P.~Braca, R.~Goldhahn, G.~Ferri, and K.~D. LePage, ``Distributed information fusion in multistatic sensor networks for underwater surveillance,'' \emph{IEEE Sensors Journal}, vol.~16, no.~11, pp. 4003--4014, Jun. 2016.

\bibitem{gao2014quantizer}
F.~Gao, L.~Guo, H.~Li, J.~Liu, and J.~Fang, ``Quantizer design for distributed {GLRT} detection of weak signal in wireless sensor networks,'' \emph{IEEE Transactions on Wireless Communications}, vol.~14, no.~4, pp. 2032--2042, Apr. 2015.

\bibitem{fang2013one}
J.~Fang, Y.~Liu, H.~Li, and S.~Li, ``One-bit quantizer design for multisensor {GLRT} fusion,'' \emph{IEEE Signal Processing Letters}, vol.~20, no.~3, pp. 257--260, Mar. 2013.

\bibitem{ciuonzo2013one}
D.~Ciuonzo, G.~Papa, G.~Romano, P.~S. Rossi, and P.~Willett, ``One-bit decentralized detection with a {Rao} test for multisensor fusion,'' \emph{IEEE Signal Processing Letters}, vol.~20, no.~9, pp. 861--864, Sep. 2013.

\end{thebibliography}
\end{document}